\documentclass[%
 reprint,
 amsmath,amssymb,
 aps,
]{revtex4-2}
\usepackage{xcolor}
\usepackage{graphicx}
\usepackage{dcolumn}
\usepackage{subcaption}
\usepackage{bm}
\usepackage{todonotes}

\begin{document}


\title{Weak-Measurement Assisted Quantum Teleportation Fidelity Enhancement under Noise}

\author{Mohit Dhanik}
\author{Shraddha Sharma}
\email{sharmas@nitrkl.ac.in, shrdha1987@gmail.com.}


\author{Pitamber Mahanandia}
\email{pitam@nitrkl.ac.in}
\affiliation{
  Department of Physics and Astronomy, NIT Rourkela, India
}%


\date{\today}

\begin{abstract}
This study presents an improved quantum teleportation protocol designed to enhance fidelity in noisy environments by combining weak measurements (WMs) with flip and reversal operations. In our scheme, Alice prepares a four-qubit entangled state and shares one of the entangled qubits with Bob, which serves as the quantum channel for teleporting an arbitrary single-qubit state. Since the communication channel is subject to noise, Alice performs a weak measurement on the shared qubit before transmission to reduce the impact of decoherence. Building upon existing WM-flip-reversal frameworks, we propose a modified weak measurement and reversal (WMR) protocol tailored for different noises in a four-qubit entangled system. The approach applies WM and flip operations prior to transmission to enhance resilience against noise, followed by corresponding reversal operations after transmission to recover the original quantum state. We systematically compare the performance of our proposed WMR protocol with the previously proposed WM-flip-reversal method under three common noise models: amplitude damping channel (ADC), phase flip channel (PFC), and bit flip channel (BFC). Our analysis reveals that the modified WMR scheme achieves significantly higher teleportation fidelity and improved robustness, particularly in bit flip noise environments. These findings highlight the potential of optimized weak measurement strategies for developing more reliable and noise-tolerant quantum communication protocols.
\end{abstract}

\maketitle


\section{\label{sec1}Introduction}
Quantum computing and quantum information science promise transformative impacts on secure communication and cryptography, as well as on computational tasks in machine learning and artificial intelligence \cite{NC2000, Pshor97, BENNETT20147, pirandola2015, biamonte2017}.
Among the key protocols enabled by quantum information principles is \textit{quantum teleportation}, where one can transfer the quantum information across separated locations, without physically transmitting the qubit {\cite{rigolin2005quantum,Furusawa1998}. Quantum  teleportation was first introduced by Bennett et al. 1993 {\cite{Bennett1993}} and soon after its proof was given by Zeiger's and popescu's group in 1997 {\cite{Bouwmeester1997}. From then, teleportation has been realized using various techniques which include photons {\cite{Mair2001,Kim2009}}, trapped ions {\cite{EnergyTeleportationIons}, atoms {\cite{Barrett2004,Riebe2004} and superconducting qubits {\cite{Kang2024}. Teleportation and the protocols related to it such as dense coding {\cite{BennettWiesner1992}}, quantum key distribution {\cite{Gisin2002}}, and remote state preparation {\cite{Bennett2000RSP}} has been studied in detail over the past three decades. {Related research directions have demonstrated teleportation protocols with built-in encryption \cite{taufiqi25} and methods to reduce the classical communication cost inherent to the teleportation procedure \cite{taufiqi25b, parakh22}, both of which represent important practical directions that motivate developing noise-robust schemes such as the one presented here.} However, teleportation has been generalized beyond qubits to higher-dimensional systems including qutrits and ququards {\cite{QuditReview}} which are majorly useful in enhancing security and increasing information capacity.
However, in practical implementations, quantum noise in the communication channel is unavoidable. Such noise leads to the degradation of entanglement and a consequent reduction in teleportation fidelity {\cite{Bouwmeester1999,Lee2000,NoiseMitigation2024,fortes2015fighting}}, thereby necessitating the development of efficient protection and error mitigation schemes. 
There are a number of methods to overcome noise effects including entanglement distillation {\cite{Bennett1996,Horodecki1996} and quantum error correction {\cite{Shor1995,Steane1996,Gottesman2009_Review} but a more resource efficient approach is to use the quantum weak measurement (WM) method {\cite{Aharonov1988,Dressel2017,harraz2020two,harraz2021quantum,li2021enhancing,harraz2022enhancing}}
which is an efficient way to protect the entangled state.
WM and its reversal have been experimentally demonstrated in superconducting qubits {\cite{Katz2008}} and photonic systems {\cite{Kim2009}} establishing viable methods for noise protection.
In this work, we focus on enhancing quantum teleportation fidelity through the use of weak measurements (WMs) combined with flip operations and their subsequent reversal. The central idea of this method is that, prior to transmitting qubits through a noisy channel, WM and flip operations are applied to improve robustness against environmental noise \cite{harraz2022enhancing}. After transmission, reversal operations are performed to recover the original state. Building on this framework, we implement the scheme on a four-qubit entangled state and propose a modified protocol that introduces new WM and reversal (WMR) operations. We then compare the resulting teleportation fidelity of our proposed protocol with that of previously proposed WM-flip-reversal scheme \cite{harraz2022enhancing}. Furthermore, we analyze the performance of both protocols under three different noise models, i.e., amplitude damping channel (ADC), phase flip channel (PFC), and bit flip channel (BFC) \cite{wu2020novel,randeep2024quantum} and demonstrate which noise environments benefit most from our approach. To the best of our knowledge, such an analysis has not been reported previously.


Having outlined the motivation for enhancing teleportation fidelity via WM and WMR operations, we now turn to a detailed description of the teleportation protocol itself. In Sec.~\ref{secii}, we present the quantum teleportation scheme incorporating weak measurements (WM) and reversal operations (WMR), which forms the basis of our proposed approach, followed by the final results in  Sec.~\ref{conc}.

\section{teleportation protocol with weak measurement}\label{secii}

It has been shown in previous works that a combination of weak measurement and flip operator can be applied to enhance the teleportation fidelity of a general single qubit state even though the quantum communication channel between Alice and Bob could be noisy \cite{harraz2022enhancing}. In this noisy teleportation, the idea is that  Alice prepares a maximally entangled quantum state and then sends one qubit to Bob through a noisy channel. So Alice and Bob share a maximally entangled state after which the usual protocol of quantum teleportation takes place. In usual teleportation protocol, in order to teleport a general single qubit to Bob, Alice makes measurements on her part of the qubits (the single qubit to be teleported and one from the maximally entangled state) and classically informs Bob of the results of her measurements which help him obtain this general state at his end. Owing to the initial maximally entangled state being shared through a noisy channel, the reduced fidelity of the state obtained by Bob at the end is evident. 

In this work, we compare two weak measurement protocols which along with the flip operators help overcome these noisy channel issues and help enhance the fidelity of obtaining a single qubit general state starting from a four qubit maximally entangled state at Alice's end, whose one of the qubit is shared to Bob through a noisy channel. We target amplitude damping, bit flip, and phase flip noises for this noisy quantum channel. 
In this section, an overview of the protocol will be given and specific forms of weak measurement and their effects will be discussed in the following subsections. Let's start from the beginning: the very initial step in this method is to start with a maximally entangled state that Alice has prepared,

\begin{eqnarray}
    |\psi_0\rangle=\frac{1}{2}\left( |0000\rangle + |0101\rangle + |1010\rangle + |1111\rangle \right)\label{initial_state}
\end{eqnarray}
{
The four-qubit state in Eq.~(\ref{initial_state}) above is chosen because it belongs to the class of generalized Bell states introduced in Refs.~\cite{Rigolin2004,rigolin2005quantum}, which are known to serve as optimal channels for deterministic quantum teleportation and support generalized multipartite Bell measurements. Unlike GHZ-type states, this state possesses maximal teleportation capability with entanglement of teleportation ($E_T=1$) \cite{rigolin2005quantum}, making it operationally distinguished for teleportation tasks. The state can further be expressed as ( $|\psi_0\rangle = |\Phi^+\rangle_{13}\otimes|\Phi^+\rangle_{24}$, where, $|\Phi^+\rangle=\frac{1}{\sqrt{2}}(|00\rangle+|11\rangle$, revealing its structure as two maximally entangled Bell pairs. Therefore, the choice of this four-qubit entangled state is also motivated by its multipartite entanglement structure, which provides enhanced robustness against local noise compared to bipartite Bell states \cite{dur2000three}.} 
}

{Once the initial state is prepared, Alice then sends the last qubit of this state to Bob through a noisy channel.} For this article, we will consider ADC, BFC, and PFC. As a result of which the qubit will face some decoherences. To overcome this decoherence a protocol by Harraz et al. \cite{harraz2022enhancing}} has been proposed for an initial two-qubit entangled state, which will be used here for our initial four-qubit entangled state along with {the new proposed protocol} for comparison. According to the protocol {in Ref.~\cite{harraz2022enhancing}}, before sending the qubit to Bob through this noisy channel, Alice will first apply the two operators 
given by $m_i^\dagger m_i,~~i=0,1$ from a family of positive operator-valued measurements (POVM) to Bob's qubit to be sent. The following subsections will discuss the exact form of these weak measurement (WM) operators. Therefore, in an initial 4 qubit entangled state, this operator takes the form $M_i=I\otimes I\otimes I\otimes m_i$, where $I$ is an identity matrix. It is to be noted that since Alice wants to send only the last qubit to Bob, the whole 4-qubit operator is an identity operator on all other qubits except the last one. This POVM is followed by flip operators with the following form,

\begin{equation}
f_0 = I = 
\begin{pmatrix}
1 & 0 \\
0 & 1
\end{pmatrix}, \quad
f_1 = \sigma_x = 
\begin{pmatrix}
0 & 1 \\
1 & 0
\end{pmatrix},\label{flip_op}
\end{equation}
where, the operation by \(f_0 (f_1)\) is decided according to the measurement outcome \(m_0 (m_1)\). The operator form for a 4-qubit system will hence be: $F_{i=0,1}=I\otimes I\otimes I\otimes f_i$. After this pair of POVM followed by pre-flip operation, the qubit to be shared with Bob will be sent through one of the noisy channels with Kraus operators $E_{i=0,1}=I\otimes I\otimes I\otimes e_i$, where $e_{i=0,1}$ is the Kraus operators corresponding to the respective noisy channel (ADC, BFC or PFC), as will be detailed in successive subsections. After sending the qubit through the noisy quantum channel, Alice uses the classical channel (to be used for teleportation) to report the measurement result 0 or 1 to Bob depending on which he would further apply post-flip ($F_i$) and reversal POVM operators or the weak measurement reversal operators (WMR) ($N_i=I\otimes I\otimes I\otimes m_i$). 

Therefore, to summarize, the whole combination would involve the sequence:\\
\begin{eqnarray}
    &~&\text{WM}(m_0/m_1)~\to \text{flip} (f_0/f_1) \to \text{channel (ADC/BFC/PFC)}\nonumber\\
    &~&\to \text{flip reversal} (f_1/f_0) \to \text{WMR} (n_0/n_1).\label{protocol}
\end{eqnarray}

\[
f_0 | 0 \rangle = | 0 \rangle
,
f_0 | 1 \rangle = | 1 \rangle
\]
\[
f_1 | 0 \rangle = | 1 \rangle
,
f_1 | 1 \rangle = | 0 \rangle
\]

In the subsequent subsections, we will take two specific WM and WMR operators and compare their results for the three damping channels, i.e. ADC, BFC, and PFC.

\subsection{Weak measurement protocol-I}\label{sec_wmi}
This protocol of WM and WMR POVM is the same protocol used in Ref.~{\cite{harraz2022enhancing}. The POVM used in this case is

\begin{equation}
m_0 = 
\begin{pmatrix}
\cos(\omega/2) & 0 \\
0 & \sin(\omega/2)
\end{pmatrix}, \quad
m_1 = 
\begin{pmatrix}
\sin(\omega/2) & 0 \\
0 & \cos(\omega/2)
\end{pmatrix},\label{adc_HZ_m}
\end{equation}
with reversal POVM operators 

\begin{equation}
n_0 = 
\begin{pmatrix}
q & 0 \\
0 & 1
\end{pmatrix}, \quad 
n_1 = 
\begin{pmatrix}
1 & 0 \\
0 & q
\end{pmatrix}.\label{adc_HZ_n}
\end{equation}

\[
m_0 | 1 \rangle = \sin(\omega/2)| 1 \rangle
,
m_0 | 0 \rangle = \cos (\omega/2) | 0 \rangle
\]
\[
m_1 | 0 \rangle = \sin (\omega/2) | 0 \rangle
,
m_1 | 1 \rangle = \cos (\omega/2 )| 1 \rangle
\]
\[
n_0 | 0 \rangle = q| 0 \rangle
,
n_0 | 1 \rangle =  | 1 \rangle
,
n_1 | 0 \rangle = | 0 \rangle
\]
\[
n_1 | 1 \rangle =  q| 1 \rangle
\].
In their published article, the authors have targeted the initial 2-qubit entangled state and targeted only the results for ADC, whereas, we would like to extend the result to a 4-qubit initial entangled state and investigate the fidelity of teleportation of single qubit considering not only ADC but also BFC and PFC as possible decoherence channels. 

\subsubsection{ADC}
In this case, the Kraus operators are:
\begin{equation}
e_0 = 
\begin{pmatrix}
1 & 0 \\
0 & \sqrt{1 - r}
\end{pmatrix}, \quad
e_1 = 
\begin{pmatrix}
0 & \sqrt{r} \\
0 & 0
\end{pmatrix},\label{ADC_e}
\end{equation}
with $r$ denoting the decaying rate within the interval $[0,1]$.
Starting from the initial state in Eq.~(\ref{initial_state}), the state that Alice and Bob share after applying the protocol summarized in Eq.~(\ref{protocol}) becomes:
\begin{eqnarray}    |\psi^{final}_{AB}\rangle&=&\frac{\sum_{i,j\in0,1} N_iF_iE_jF_iM_i|\psi_0\rangle}{\sqrt{\sum_{i,j\in0,1}\parallel N_iF_iE_jF_iM_i|\psi_0\rangle\parallel^2}}\label{psi_final_adcI}
\end{eqnarray}

Using Eq.~(\ref{initial_state}), (\ref{flip_op}), (\ref{adc_HZ_m}), (\ref{adc_HZ_n}) in Eq.~(\ref{psi_final_adcI}), we obtain the final state as:
\begin{eqnarray}    |\psi^{final}_{AB}\rangle&=&\lambda_{ADC}(  |0000\rangle+|1010\rangle+|0101\rangle+|1111\rangle )\nonumber\\
&+&(|0100\rangle+|1110\rangle+|0001\rangle+|1011\rangle )
\end{eqnarray}

\noindent where, $\lambda_{ADC}=\frac{\cos{(\omega/2)}q+\sin{(\omega/2)}\sqrt{1-r}}{\sqrt{2(\cos^2{\omega/2}q^2+\sin^2{\omega/2}(1-r))}}$
$r$.
\begin{figure*}[ht]
  \centering
  \begin{minipage}[b]{0.4\textwidth}    \includegraphics[width=1\linewidth]{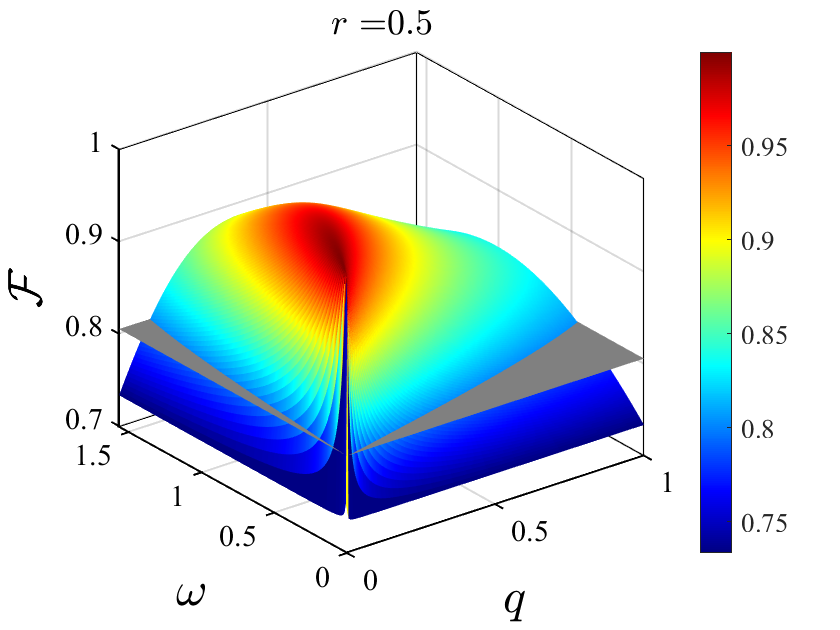}
  \end{minipage}
  \hfill
  \begin{minipage}[b]{0.4\textwidth}    \includegraphics[width=1\linewidth]{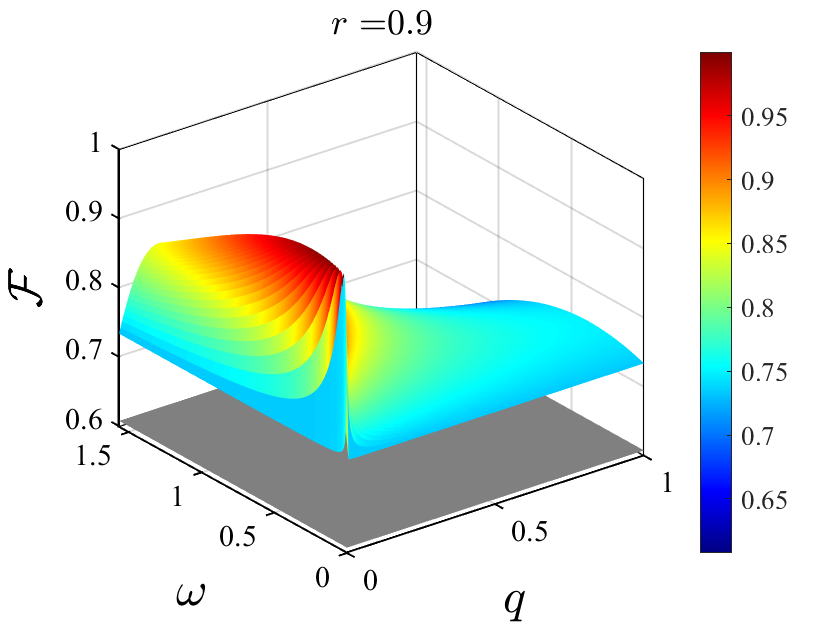}
  \end{minipage}
  \caption{{Teleportation fidelity for ADC with weak measurement protocol-I presented in Sec.~(\ref{sec_wmi}). We compare ADC for $r=0.5$ (left panel) and $r=0.9$ (right panel) and observe that the protocol provides a higher fidelity than the unprotected teleportation protocol (shown by the gray region) for a range of parameter values. }}\label{ADC_sHWM}
\end{figure*}

In the state $|\psi^{final}_{AB}\rangle$, the first three qubits are now with Alice, and the last qubit is with Bob, i.e., $|AAAB\rangle$ is the way the qubits in this 4-qubit state are distributed. After this, Alice incorporates the input qubit in the weak measurement protected state, $|\psi^{final}_{AB}\rangle$. The combined state then takes the form:

\begin{eqnarray}    |\psi_{in,A,B}^{ADC,~I}\rangle&=&|\psi_{in}\rangle\otimes|\psi_{AB}^{final}\rangle\\
&=&(\alpha|0\rangle+\beta|1\rangle)\otimes|\psi_{AB}^{final}\rangle\nonumber\\
&=&\alpha(\lambda_{ADC}(|0000\rangle+|0101\rangle)+|0010\rangle+|0111\rangle   )|0\rangle\nonumber\\
&+&\alpha(\lambda_{ADC}(|0010\rangle+|0111\rangle)+|0000\rangle+|0101\rangle   )|1\rangle\nonumber\\
&+&\beta(\lambda_{ADC}(|1000\rangle+|1101\rangle)+|1010\rangle+|1111\rangle   )|0\rangle\nonumber\\
&+&\beta(\lambda_{ADC}(|1010\rangle+|1111\rangle)+|1000\rangle+|1101\rangle )|1\rangle\nonumber\\
\end{eqnarray}{}
\noindent The above state can be rearranged as:
\begin{eqnarray}    |\psi_{in,A,B}^{ADC,I}\rangle&=& \eta_1\lambda_{ADC}(\alpha|0\rangle+\beta|1\rangle)+\eta_2\lambda_{ADC}(\alpha|0\rangle-\beta|1\rangle)\nonumber\\    &+&\eta_1(\alpha|0\rangle+\beta|1\rangle)+\eta_2(\alpha|1\rangle-\beta|0\rangle)\nonumber\\
    &+&\eta_3\lambda_{ADC}(\alpha|1\rangle+\beta|0\rangle)+\eta_4\lambda_{ADC}(\alpha|1\rangle-\beta|0\rangle)\nonumber\\     &+&\eta_3(\alpha|1\rangle+\beta|0\rangle)+\eta_4(\alpha|1\rangle-\beta|0\rangle)
\end{eqnarray}
\noindent where, 
\begin{eqnarray}
    \eta_{1,2}&=&=\frac{1}{2}\left[\left(|0000\rangle+|0101\rangle\right)\pm\left(|1010\rangle+|1111\rangle \right)\right],~\text{and,}\nonumber\\
    \eta_{3,4}&=&=\frac{1}{2}\left[\left(|0010\rangle+|0111\rangle\right)\pm\left(|1000\rangle+|1101\rangle \right)\right]
\end{eqnarray}

 In these $\eta$'s, the qubits are arranged as $|inAAA\rangle$, \textit{i.e.}, the first qubit is the input qubit, whereas the last 3 qubits are those that Alice holds from the shared entangled state with Bob. Whereas, the qubit appearing with $\alpha$ and $\beta$ is the qubit that Bob has, which was given to him by Alice after the weak measurement protection protocol, before she included the input qubit. These $\eta_{1,2,3,4}$ provide a new basis for the projection operation to 4 out of 16 generalized Bell states \cite{rigolin2005quantum}, or G-states as $\mathcal{B}_i=|\eta_{i}\rangle\langle \eta_{i}|$, where $i\in 1,2,3,4$. 

Once Alice has performed the projective measurement, she conveys this information $i$, in other words, which projective measurement she has performed, to Bob through a classical channel. Therefore, the state that Bob is left with then becomes:

\begin{eqnarray}
    \rho_{{B}_i}=Tr_{inAAA} (\mathcal{B}_i|\psi_{in,A,B}^{ADC,I}\rangle \langle\psi_{in,A,B}^{ADC,I}|\mathcal{B}_i^\dagger),
\end{eqnarray}

\noindent where $Tr_{inAAA}$ is the trace over all Alice's qubits (input and her part of the entangled state). It should be noted that the above state is not normalized. Afterwards, based on the one-qubit information from Alice, Bob will apply the corresponding reversing operator to the state he has. Table \ref{ADC_SS_table} summarizes the operators corresponding to the projection operator.

\begin{eqnarray}    \rho_{\mathcal{ R}_i}=\frac{\mathcal{R}_i\rho_{{B}_i}\mathcal{R}_i^\dagger}{Tr(\rho_{B_i})}
\end{eqnarray}

\begin{table*}[ht]
\caption{\label{ADC_SS_table}}
\begin{ruledtabular}
\begin{tabular}{ccc}
 Alice's outcome&Bob's outcome&Bob's unitary operation\\ \hline
 $|\eta_1\rangle=\frac{1}{2}\left(|0000\rangle+|0101\rangle+|1010\rangle+|1111\rangle \right)$&$\alpha|0\rangle+\beta|1\rangle$ & $I$ \\
 $|\eta_2\rangle=\frac{1}{2}\left(|0000\rangle+|0101\rangle-|1010\rangle-|1111\rangle \right)$&$\alpha|0\rangle-\beta|1\rangle$ & $\sigma_z$ \\
 $|\eta_3\rangle=\frac{1}{2}\left(|0010\rangle+|0111\rangle+|1000\rangle+|1101\rangle \right)$&$\alpha|1\rangle+\beta|0\rangle$ & $\sigma_x$ \\
 $|\eta_4\rangle=\frac{1}{2}\left(|0010\rangle+|0111\rangle-|1000\rangle-|1101\rangle \right)$&$\alpha|1\rangle-\beta|0\rangle$ & $\sigma_z\sigma_x$\\
\end{tabular}
\end{ruledtabular}
\end{table*}
The fidelity, therefore, is the overlap of Bob's state obtained after applying the reversing operator and the input qubit state that Alice wanted to teleport. The expression, therefore, takes the form:

\begin{eqnarray}
    \mathcal{F}=\sum_{i=1}^4 Tr(\rho_{{B}_i})\langle \psi_{in}|\rho_{\mathcal{R}_i}|\psi_{in}\rangle.
\end{eqnarray}
\noindent However, it can be observed that the above fidelity depends on $\alpha$ and $\beta$. Consequently, the final fidelity is averaged over all possible input parameters, in order to make it independent of any specific input state

\begin{eqnarray}
    \mathcal{F}=\int d\psi_{in} \sum_{i=1}^4 Tr(\rho_{{B}_i})\langle \psi_{in}|\rho_{\mathcal{R}_i}|\psi_{in}\rangle.
\end{eqnarray}

{The behavior of $\mathcal{F}$ is shown in Fig.~(\ref{ADC_sHWM}) where the gray region is the fidelity calculated without any weak measurement protection, and an enhancement of fidelity can clearly be observed after the WM protocol-I is implemented even as the decoherence is increased}.


\subsubsection{BFC}
{Let us now explore the efficiency of this protocol for bit-flip decoherence channel}. For BFC, the Kraus operators are: 
\begin{equation}
e_0 = 
\sqrt{1-r}\begin{pmatrix}
1 & 0 \\
0 & 1
\end{pmatrix}, \quad
e_1 = 
\sqrt{r}\begin{pmatrix}
0 & 1 \\
1 & 0
\end{pmatrix},\label{bfc_e}
\end{equation}

\begin{figure}[h]
  \centering
  \begin{minipage}[b]{0.4\textwidth}    \includegraphics[width=1\linewidth]{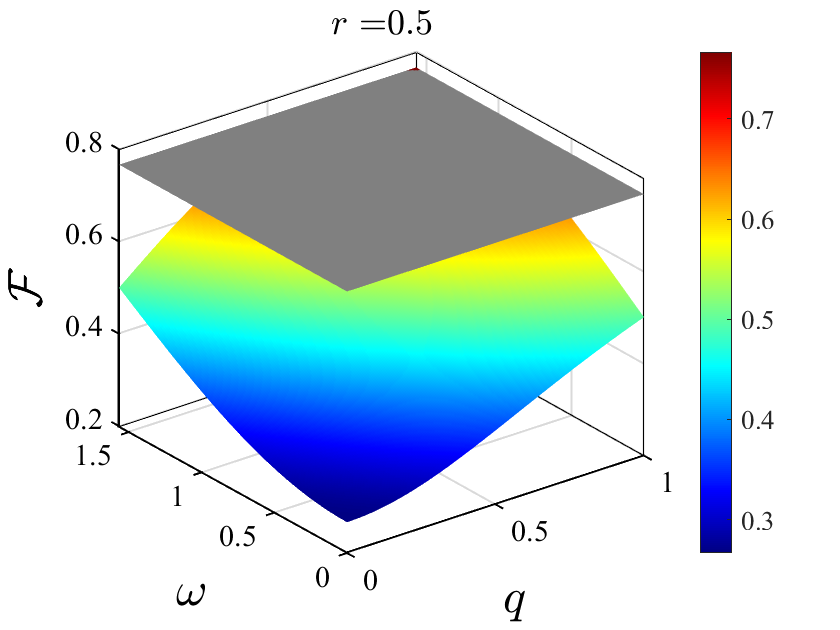}
  \end{minipage}
  \hfill
  \begin{minipage}[b]{0.4\textwidth}    \includegraphics[width=1\linewidth]{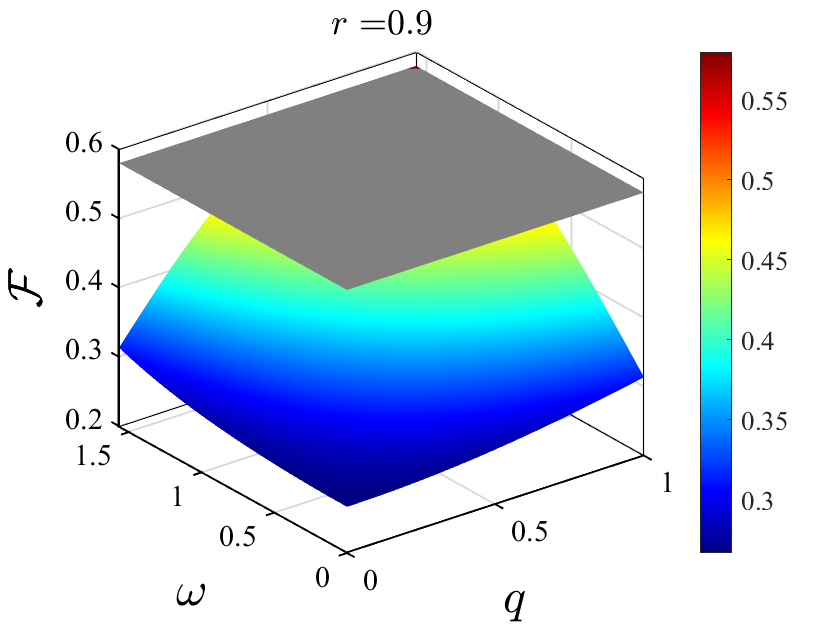}
  \end{minipage}
  \caption{{Teleportation fidelity for the BFC under WM Protocol-I, as described in Sec.~(\ref{sec_wmi}). The fidelity is shown for decoherence strengths $r=0.5$ (top) and $r=0.9$ (bottom). The protocol does not yield higher teleportation fidelity than the unprotected protocol (gray region) over the whole range of parameter values.}}\label{BFC_sHWM}
\end{figure}
 Now using operators from Eq.~(\ref{adc_HZ_m}) followed by Eq.~(\ref{bfc_e}) and Eq.~(\ref{adc_HZ_n}) and operating on Eq.~(\ref{initial_state}), we obtain the final state Eq.~(\ref{psi_final_adcI}) as:

\begin{eqnarray}    |\psi^{final}_{AB}\rangle&=&\lambda_{BFC_1}(  |0000\rangle+|1010\rangle+|0101\rangle+|1111\rangle )\nonumber\\
&+
&\lambda_{BFC_2}(|0100\rangle+|1110\rangle+|0001\rangle+|1011\rangle)
\end{eqnarray}
\noindent where, $\lambda_{BFC_1}=\frac{1}{2} [ (1-p) \sin^2 \omega/2 + q^2 (1-p) \cos^2 \omega/2]$\\

and, $\lambda_{BFC_2}=\frac{1}{2} [ pq^2 \sin^2 \omega/2 + p \cos^2 \omega/2]$.

{In the state $|\psi^{final}_{AB}\rangle$, the first three qubits are now with Alice and the last qubit is with Bob.} After this, Alice interacts her qubit, which she wants to send  with the protected entangled state, $|\psi^{final}_{AB}\rangle$. Then the combined state  takes the form:

\begin{eqnarray}    |\psi_{in,A,B}^{BFC,~I}\rangle&=&|\psi_{in}\rangle\otimes|\psi_{AB}^{final}\rangle\\
&=&(\alpha|0\rangle+\beta|1\rangle)\otimes|\psi_{AB}^{final}\rangle\nonumber\\
&=&\alpha[\lambda_{BFC_1}(|0000\rangle+|0101\rangle)|0\rangle\nonumber+\\&&\lambda_{BFC_1}(|0010\rangle+|0111\rangle   )|1\rangle\nonumber+\\
&&\lambda_{BFC_2}(|0010\rangle+|0111\rangle)|0\rangle\nonumber+\\&&\lambda_{BFC_2}(|0000\rangle+|0101\rangle   |1\rangle\nonumber]+\\
&&\beta[\lambda_{BFC_1}(|1000\rangle+|1101\rangle)|0\rangle\nonumber+\\&&\lambda_{BFC_1}(|1010\rangle+|1111\rangle   )|1\rangle\nonumber+\\
&&\lambda_{BFC_2}(|1010\rangle+|1111\rangle)|0\rangle\nonumber+\\&&\lambda_{BFC_2}(|1000\rangle+|1101\rangle   |1\rangle\nonumber]\\
\end{eqnarray}{}
\noindent The above state can be rearranged as:
{\small{
\begin{eqnarray}    |\psi_{in,A,B}^{BFC,I}\rangle&=& \eta_1(\alpha|0\rangle+\beta|1\rangle)+\eta_2(\alpha|0\rangle-\beta|1\rangle)\nonumber\\   
    &+&\eta_3(\alpha|1\rangle+\beta|0\rangle)+\eta_4(\alpha|1\rangle-\beta|0\rangle)\nonumber\\     
\end{eqnarray}
\noindent where, 
\begin{eqnarray}
    \eta_{1,2}&=&\frac{1}{2}\bigg[[\lambda_{BFC_1}(|0000\rangle+|0101\rangle)\nonumber+\lambda_{BFC_2}(|0010\rangle+|0111\rangle   )\nonumber]\\
   &&\pm [\lambda_{BFC_1}(|1010\rangle+|1111\rangle)|\nonumber+\lambda_{BFC_2}(|1000\rangle+|1101\rangle\nonumber]\bigg]
 \end{eqnarray} 
 \begin{eqnarray}
    \eta_{3,4}&=&\frac{1}{2}\bigg[[\lambda_{BFC_1}(|0010\rangle+|0111\rangle)\nonumber+\lambda_{BFC_2}(|0000\rangle+|0101\rangle   )\nonumber]\\
   &&\pm [\lambda_{BFC_1}(|1000\rangle+|1101\rangle)|\nonumber+\lambda_{BFC_2}(|1010\rangle+|1111\rangle\nonumber]\bigg]\\
\end{eqnarray}
}}

{ Following the same steps as in the case of ADC, we obtain the fidelity in the case of BFC decoherence as shown in Fig.~\ref{BFC_sHWM}. It can be observed from these plots that utilizing WM protocol-I shows no advantage over the unprotected fidelity values. Therefore, this protocol fails to obtain higher fidelity in the case of BFC decoherence.}

\subsubsection{PFC}

{We extend this analysis of WM protection protocol-I to the PFC decoherence}. The Kraus operators associated with PFC are:

\begin{equation}
e_0 = 
\sqrt{1-r}\begin{pmatrix}
1 & 0 \\
0 & 1
\end{pmatrix}, \quad
e_1 = 
\sqrt{r}\begin{pmatrix}
1 & 0 \\
0 & -1
\end{pmatrix},\label{pfc_e}
\end{equation}
where $r$ is the decoherence strength.
\begin{figure}[h]
  \centering
  \begin{minipage}[b]{0.4\textwidth}    \includegraphics[width=1\linewidth]{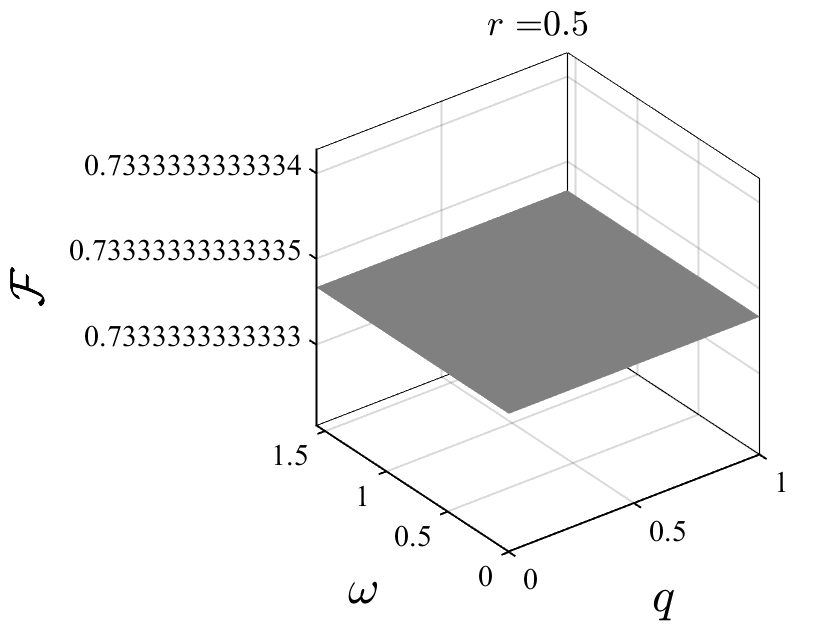}
  \end{minipage}
  \hfill
  \begin{minipage}[b]{0.4\textwidth}    \includegraphics[width=1\linewidth]{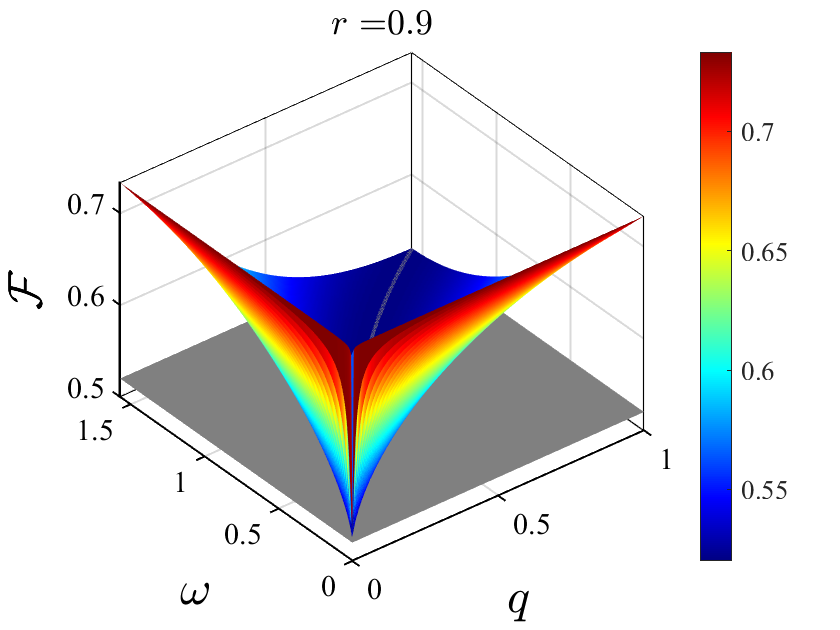}
  \end{minipage}
  \caption{{Teleportation fidelity for PFC with WM protocol-I for $r=0.5$ (top) and $r=0.9$ (bottom) values. In the case of $r=0.5$, the fidelity is same as the unprotected teleportation protocol (shown by the gray region), but for $r=0.9$ the fidelity is higher for a range of parameter values.}}\label{PFC_sHWM}
\end{figure}
Now, following the similar steps as followed for ADC and BFC, we obtain the final state as:
 \begin{eqnarray}    |\psi^{final}_{AB}\rangle&=&\lambda_{PFC}(  |0000\rangle+|1010\rangle+|0101\rangle+|1111\rangle )\nonumber
\end{eqnarray}
\noindent where, $\lambda_{PFC}=\frac{q\cos{(\omega/2)}{}}{2\sqrt{q^2\cos^2{\omega/2}+\sin^2{\omega/2}}}$

The first three qubits of  $|\psi^{final}_{AB}\rangle$ belongs to Alice and the last qubit to Bob. After this, Alice interacts her qubit  with the  protected entangled state, $|\psi^{final}_{AB}\rangle$. Then the combined state  takes the form:
\begin{eqnarray}    |\psi_{in,A,B}^{PFC,~I}\rangle&=&|\psi_{in}\rangle\otimes|\psi_{AB}^{final}\rangle\\
&=&(\alpha|0\rangle+\beta|1\rangle)\otimes|\psi_{AB}^{final}\rangle\nonumber\\
&=&\alpha(\lambda_{PFC}(|0000\rangle+|0101\rangle)|0\rangle\nonumber\\
&+&\alpha(\lambda_{PFC}(|0010\rangle+|0111\rangle)|1\rangle\nonumber\\
&+&\beta(\lambda_{PFC}(|1000\rangle+|1101\rangle)|0\rangle\nonumber\\
&+&\beta(\lambda_{PFC}(|1010\rangle+|1111\rangle)|1\rangle\nonumber\\
\end{eqnarray}{}
\noindent The above state can be rearranged as:
\begin{eqnarray}    |\psi_{in,A,B}^{PFC,I}\rangle&=& \eta_1\lambda_{PFC}(\alpha|0\rangle+\beta|1\rangle)+\eta_2\lambda_{PFC}(\alpha|0\rangle-\beta|1\rangle)\nonumber\\    &+&\eta_1(\alpha|0\rangle+\beta|1\rangle)+\eta_2(\alpha|1\rangle-\beta|0\rangle)\nonumber\\
    &+&\eta_3\lambda_{PFC}(\alpha|1\rangle+\beta|0\rangle)+\eta_4\lambda_{PFC}(\alpha|1\rangle-\beta|0\rangle)\nonumber\\     &+&\eta_3(\alpha|1\rangle+\beta|0\rangle)+\eta_4(\alpha|1\rangle-\beta|0\rangle)\label{pfc_psi}
\end{eqnarray}
\noindent where, 
\begin{eqnarray}
    \eta_{1,2}&=&=\frac{1}{2}\left[\left(|0000\rangle+|0101\rangle\right)\pm\left(|1010\rangle+|1111\rangle \right)\right],~\text{and,}\nonumber\\
    \eta_{3,4}&=&=\frac{1}{2}\left[\left(|0010\rangle+|0111\rangle\right)\pm\left(|1000\rangle+|1101\rangle \right)\right]
\end{eqnarray}

{The behavior of $\mathcal{F}$ for this decoherence utilizing the state in Eq.~(\ref{pfc_psi}), we observe that as $r$ is increased, the WM protocol-I starts to perform better when compared to the case of teleportation without any protection. Therefore, WM protocol-I works well for PFC decoherence for larger $r$ values.}

{Having presented a comprehensive analysis of the performance of WM Protocol-I, we now introduce a new WM scheme, Protocol-II, and investigate the behavior of the teleportation fidelity under ADC, BFC, and PFC decoherence channels.}

\subsection{Weak measurement protocol-II}\label{sec_wmii}
{In view of the observation that WM Protocol-I performs well only in the presence of ADC noise and, for PFC noise, only at relatively large values of $r$, we introduce WM Protocol-II to improve the teleportation fidelity under BFC while preserving its effectiveness for the ADC and PFC channels. The motivation, intuition and conceptual differences of the operators in this new protocol has been discussed in Appendix~(\ref{app1}). The operators for WM protocol-II are:}
\begin{equation}
m_0 = 
\begin{pmatrix}
K_1^+ & 0 \\
0 & K_1^-
\end{pmatrix}, \quad
m_1 = 
\begin{pmatrix}
K_1^- & 0 \\
0 & K_1^+
\end{pmatrix},\label{new_WM_POVM}
\end{equation}
where, $K_1^+=\sqrt{\frac{1+K_1}{2}}$, $K_1^-=\sqrt{\frac{1-K_1}{2}}$, with reversal POVM operators:

\begin{equation}
n_0 = 
\begin{pmatrix}
K_2^+ & 0 \\
0 & K_2^-
\end{pmatrix}, \quad 
n_1 = 
\begin{pmatrix}
K_2^- & 0 \\
0 & K_2^+
\end{pmatrix}\label{new_WMR_POVM},
\end{equation}
here, $K_2^+=\sqrt{\frac{1+K_2}{2}}$ and, $K_2^-=\sqrt{\frac{1-K_2}{2}}$.\\

{
The weak measurement operators used in Protocol-II are constructed using parameters $K_1$ and $K_2$, which allow a symmetric and tunable control over measurement strength. Unlike Protocol-I, where the measurement strength is governed by trigonometric parameters, Protocol-II rewrites it in parameters such that $K_{1/2}\in[-1,1]$. On clear comparison with WMR operators in Eq.~(\ref{adc_HZ_n}) of Protocol-I, the operators in Protocol-II of Eq.~(\ref{new_WMR_POVM}) are more symmetric to account of not just spontaneous decay (i.e., ADC-like noise) but also for decoherence like BFC. The corresponding reversal operators are designed to probabilistically invert the disturbance introduced by the weak measurement. This enables partial recovery of the quantum state after it has interacted with the noisy environment. The effectiveness of this protocol depends on the balance between measurement strength and reversibility, which determines the overall fidelity and success probability.
}
\subsubsection{ADC}
The ADC will be implemented using Kraus operators in Eq.~(\ref{ADC_e}). We will be applying again {Eq.~(\ref{psi_final_adcI})}, for this case, we obtain:
\begin{eqnarray}
|\psi^{final,~ADC}_{AB}\rangle&=&\lambda_1( |0000\rangle +|0101\rangle+|1010\rangle+|1111\rangle)\nonumber\\
&~&+|0100\rangle+|1110\rangle+|0001\rangle+|1011\rangle,\nonumber\\ 
\end{eqnarray}
\noindent where, $\lambda_1=\frac{\sqrt{2}\left( \sqrt{K_1^+K_2^+}+\sqrt{(K_1^-K_2^-)(1-r)}\right)}{\sqrt{(K_1^+K_2^+)^2+(1-r)(K_1^-K_2^-)^2}}$.

Afterward, as per the usual teleportation protocol, Alice brings in the input qubit to be teleported. With this qubit being in a general state $|\psi_{in}\rangle=\alpha|0\rangle+\beta|1\rangle$, the state of the joint system becomes:

\begin{eqnarray}    |\psi_{in,A,B}^{ADC,~II}\rangle&=&|\psi_{in}\rangle\otimes|\psi^{final,~ADC}_{AB}\rangle\\
&=&(\alpha|0\rangle+\beta|1\rangle)\otimes|\psi^{final,~ADC}_{AB}\rangle\nonumber\\
&=&\alpha(\lambda_{1}(|0000\rangle+|0101\rangle)+|0010\rangle+|0111\rangle   )|0\rangle\nonumber\\
&+&\alpha(\lambda_{1}(|0010\rangle+|0111\rangle)+|0000\rangle+|0101\rangle   )|1\rangle\nonumber\\
&+&\beta(\lambda_{1}(|1000\rangle+|1101\rangle)+|1010\rangle+|1111\rangle   )|0\rangle\nonumber\\
&+&\beta(\lambda_{1}(|1010\rangle+|1111\rangle)+|1000\rangle+|1101\rangle )|1\rangle\nonumber\\
\end{eqnarray}{}
\noindent The above state can be rearranged as:
\begin{eqnarray}    |\psi_{in,A,B}^{ADC,II}\rangle&=& \eta_1\lambda_{1}(\alpha|0\rangle+\beta|1\rangle)+\eta_2\lambda_{1}(\alpha|0\rangle-\beta|1\rangle)\nonumber\\    &+&\eta_1(\alpha|0\rangle+\beta|1\rangle)+\eta_2(\alpha|1\rangle-\beta|0\rangle)\nonumber\\
    &+&\eta_3\lambda_{1}(\alpha|1\rangle+\beta|0\rangle)+\eta_4\lambda_{1}(\alpha|1\rangle-\beta|0\rangle)\nonumber\\     &+&\eta_3(\alpha|1\rangle+\beta|0\rangle)+\eta_4(\alpha|1\rangle-\beta|0\rangle)
\end{eqnarray}
\noindent where, 
\begin{eqnarray}
    \eta_{1,2}&=&=\frac{1}{2}\left[\left(|0000\rangle+|0101\rangle\right)\pm\left(|1010\rangle+|1111\rangle \right)\right],~\text{and,}\nonumber\\
    \eta_{3,4}&=&=\frac{1}{2}\left[\left(|0010\rangle+|0111\rangle\right)\pm\left(|1000\rangle+|1101\rangle \right)\right]
\end{eqnarray}
Subsequently, Alice makes a projective measurement on her part of the qubits, i.e. 1 input qubit and 3 qubits out of the 4-qubit entangled state she shared with Bob. Therefore, in state $|\psi^{ADC}_{in, A, B}\rangle$, the first 4 qubits belong to Alice, whereas, the last qubit is with Bob. Succeeding the joint measurement on input qubits and her part of the 3 qubits, Alice sends Bob a 1-bit classical message informing him which state she has measured. This crucial information helps Bob apply the corresponding unitary operator to recover the single qubit to be teleported faithfully. In Table~(\ref{ADC_SS_table}), we present the projective measurements by Alice and the corresponding unitary operations for Bob.

In Fig.~(\ref{ADC_newWM}),
we show the behavior of teleportation fidelity for different values of the decaying rate $r$. {It can be observed that the WM protocol-II also display higher fidelity when compared to the case of unprotected teleportation fro certain range of parameter values.}
\begin{figure*}[ht]
  \centering
  \begin{minipage}[b]{0.4\textwidth}    \includegraphics[width=1\linewidth]{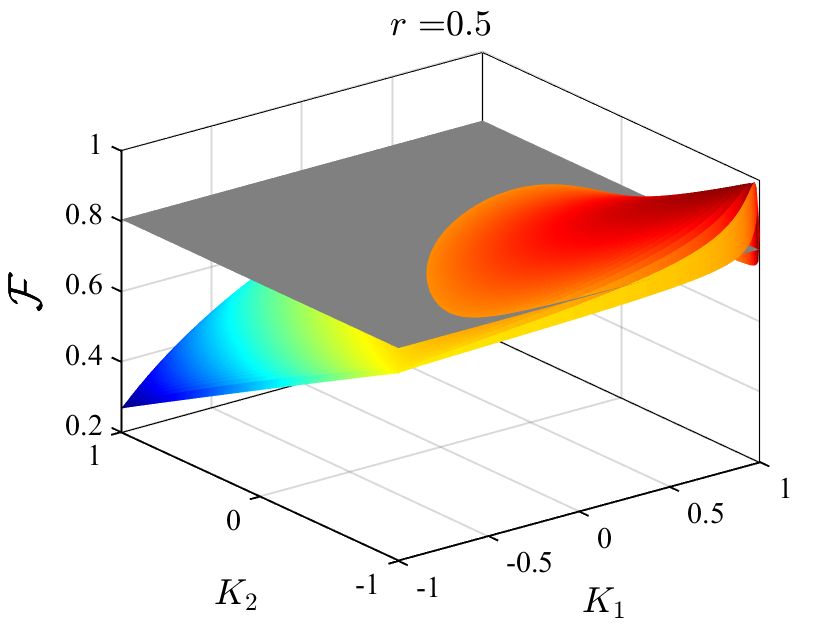}
  \end{minipage}
  \hfill
  \begin{minipage}[b]{0.4\textwidth}    \includegraphics[width=1\linewidth]{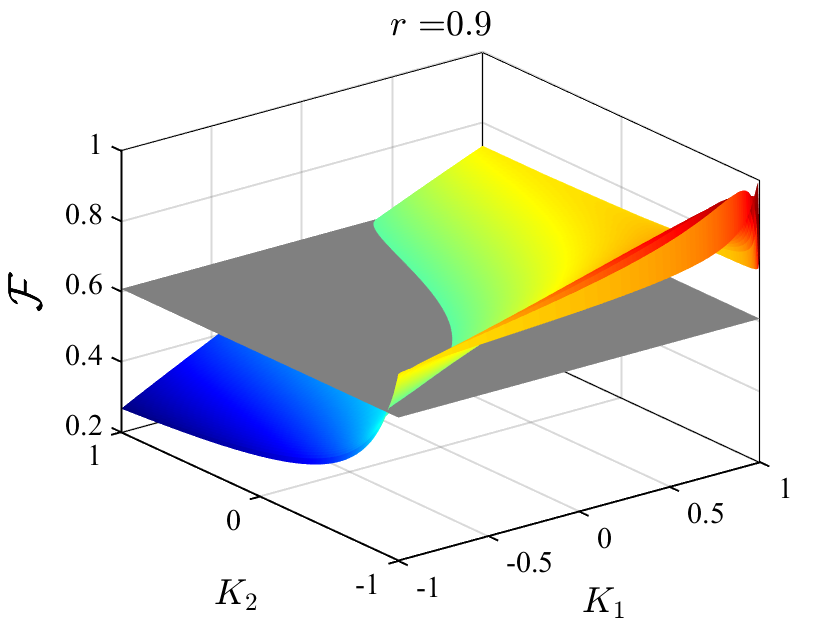}
  \end{minipage}
  \caption{Teleportation fidelity for ADC with WM protocol-II. {We compare ADC for $r=0.5$ (left panel) and $r=0.9$ (right panel) and observe that the protocol provides a higher fidelity than the unprotected teleportation protocol (shown by the gray region) for a range of parameter values.}}\label{ADC_newWM}
\end{figure*}
\begin{figure*}[ht]
  \centering
  \begin{minipage}[b]{0.4\textwidth}    \includegraphics[width=1\linewidth]{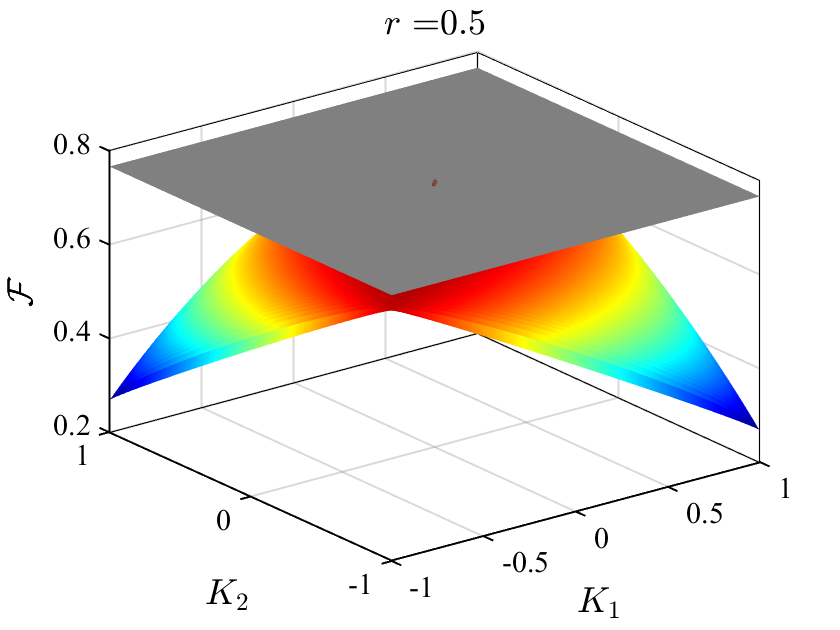}
  \end{minipage}
  \hfill
  \begin{minipage}[b]{0.4\textwidth}    \includegraphics[width=1\linewidth]{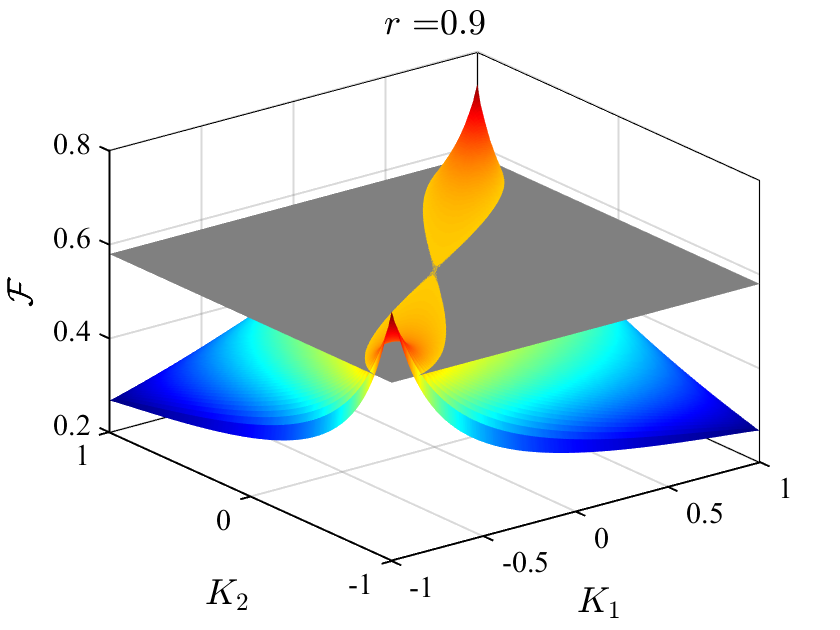}
  \end{minipage}
  \caption{{Teleportation fidelity for the BFC with WM Protocol-II. Results are shown for decoherence strengths $r=0.5$ (left panel) and $r=0.9$ (right panel). For $r=0.5$, the protocol yields teleportation fidelity comparable to that of the unprotected protocol (gray region), whereas for $r=0.9$, it achieves higher fidelity than the unprotected protocol over a range of parameter values.}}\label{BFC_newWM}
\end{figure*}

\subsubsection{BFC}

]textcolor{blue}{For the case of BFC,} using Eq.~(\ref{initial_state}), (\ref{flip_op}), (\ref{bfc_e}), (\ref{new_WM_POVM}) and (\ref{new_WMR_POVM}), the state after Alice has sent 4th qubit to Bob in this case becomes:
\begin{figure*}[ht]
  \centering
  \begin{minipage}[b]{0.4\textwidth}    \includegraphics[width=1\linewidth]{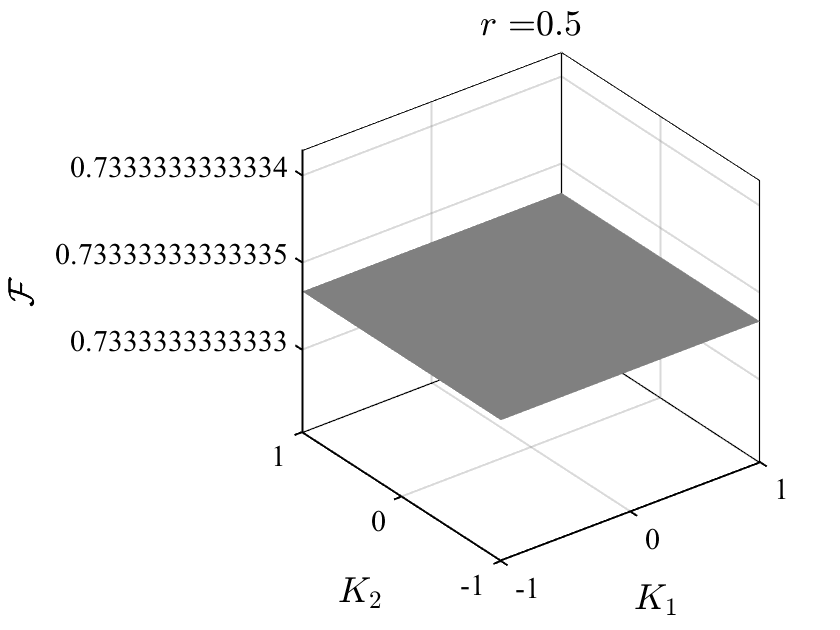}
  \end{minipage}
  \hfill
  \begin{minipage}[b]{0.4\textwidth}    \includegraphics[width=1\linewidth]{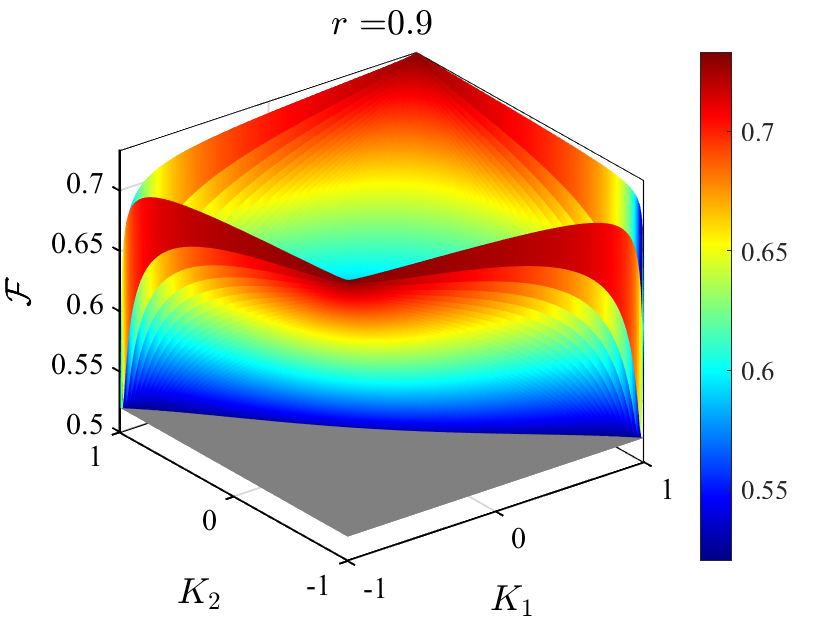}
  \end{minipage}
  \caption{{Teleportation fidelity for PFC with weak measurement protocol-II. {We compare PFC for $r=0.5$ (left panel) and $r=0.9$ (right panel) and observe that for $r=0.5$  the protocol provides a same fidelity as for the unprotected teleportation protocol(shown by the gray region) but for $r=0.9$ it provides higher fidelity than the unprotected protocol for a range of parameter values.}}{
  }}\label{PFC_newWM}
\end{figure*}
\begin{eqnarray}    |\psi^{final,~BFC}_{AB}\rangle&=&\lambda_1( |0000\rangle +|0101\rangle+|1010\rangle+|1111\rangle)\nonumber\\
&~&+\lambda_2(|0100\rangle+|1110\rangle+|0001\rangle+|1011\rangle),\nonumber\\ \label{psi_new_bfc} 
\end{eqnarray}
where, 
\begin{eqnarray}
   \lambda_1&=&\frac{K_1^+K_2^++K_1^-K_2^-}{\sqrt{2\left((K_1^+K_2^+)^2+(K_1^-K_2^-)^2 \right)}}, \nonumber\\
   \lambda_2&=&\frac{K_1^+K_2^-+K_1^-K_2^+}{\sqrt{2\left((K_1^+K_2^-)^2+(K_1^-K_2^+)^2 \right)}}.
\end{eqnarray}

Alice then brings the input qubit to connect it to her three existing qubits, leading to:
\begin{eqnarray}    |\psi_{in,A,B}^{BFC,~II}\rangle&=&|\psi_{in}\rangle\otimes|\psi^{final,~BFC}_{AB}\rangle\\
&=&(\alpha|0\rangle+\beta|1\rangle)\otimes|\psi^{final,~BFC}_{AB}\rangle\nonumber\\
&=&\alpha[\lambda_{1}(|0000\rangle+|0101\rangle)|0\rangle\nonumber+\\&&\lambda_{1}(|0010\rangle+|0111\rangle   )|1\rangle\nonumber+\\
&&\lambda_{2}(|0010\rangle+|0111\rangle)|0\rangle\nonumber+\\&&\lambda_{2}(|0000\rangle+|0101\rangle   |1\rangle\nonumber]+\\
&&\beta[\lambda_{1}(|1000\rangle+|1101\rangle)|0\rangle\nonumber+\\&&\lambda_{1}(|1010\rangle+|1111\rangle   )|1\rangle\nonumber+\\
&&\lambda_{2}(|1010\rangle+|1111\rangle)|0\rangle\nonumber+\\&&\lambda_{2}(|1000\rangle+|1101\rangle   |1\rangle\nonumber]\\
\end{eqnarray}{}
\noindent The above state can be rearranged as:
\begin{eqnarray}    |\psi_{in,A,B}^{BFC,II}\rangle&=& \eta_1(\alpha|0\rangle+\beta|1\rangle)+\eta_2(\alpha|0\rangle-\beta|1\rangle)\nonumber\\   
    &+&\eta_3(\alpha|1\rangle+\beta|0\rangle)+\eta_4(\alpha|1\rangle-\beta|0\rangle)\nonumber\\     
\end{eqnarray}

\noindent where, 
\begin{eqnarray}
    \eta_{1,2}&=&\frac{1}{2}\bigg[[\lambda_{1}(|0000\rangle+|0101\rangle)\nonumber+\lambda_{2}(|0010\rangle+|0111\rangle   )\nonumber]\\
   &&\pm [\lambda_{1}(|1010\rangle+|1111\rangle)|\nonumber+\lambda_{2}(|1000\rangle+|1101\rangle\nonumber]\bigg]
 \end{eqnarray} 
 \begin{eqnarray}
    \eta_{3,4}&=&\frac{1}{2}\bigg[[\lambda_{1}(|0010\rangle+|0111\rangle)\nonumber+\lambda_{2}(|0000\rangle+|0101\rangle   )\nonumber]\\
   &&\pm [\lambda_{1}(|1000\rangle+|1101\rangle)|\nonumber+\lambda_{2}(|1010\rangle+|1111\rangle\nonumber]\bigg]\\
\end{eqnarray}

In this case also,  the measurements that Alice performs and the corresponding unitary operations Bob must perform in order to create the qubit at his end are given in Table~\ref{ADC_SS_table}. The resulting fidelity is displayed in Fig.~(\ref{BFC_newWM}) { which unlike the case of WM protocol-I exhibits higher fidelity value compared to that of unprotected  teleportation via BFC as $r$ is increased}.

\subsubsection{PFC}
{Following the same procedure as in the previous cases, we apply the Kraus operators in Eq.~(\ref{pfc_e}), together with the POVM and flip operators of WM Protocol-II, to the initial state in Eq.~(\ref{initial_state}) in the order specified by Eq.~(\ref{protocol}). The resulting shared state between Alice and Bob after the complete protocol is}

\begin{eqnarray}
    |\psi_{AB}^{final,PFC}\rangle=\lambda\left( |0000\rangle+|1010\rangle+|0101\rangle+|1111\rangle  \right),\nonumber\\
\end{eqnarray}
where, $\lambda=k_1^+k_2^+\sqrt{{2}/({(k_1^-k_2^-)^2+(k_1^+k_2^+)^2}})$.\\
Alice then interacts with her qubit with the protected entangled state. Hence, the combined state becomes:
\begin{eqnarray}    |\psi_{in,A,B}^{PFC,~II}\rangle&=&|\psi_{in}\rangle\otimes|\psi^{final,~PFC}_{AB}\rangle\\
&=&(\alpha|0\rangle+\beta|1\rangle)\otimes|\psi^{final,~PFC}_{AB}\rangle\nonumber\\
&=&\alpha(\lambda_{}(|0000\rangle+|0101\rangle)|0\rangle\nonumber\\
&+&\alpha(\lambda_{}(|0010\rangle+|0111\rangle)|1\rangle\nonumber\\
&+&\beta(\lambda_{}(|1000\rangle+|1101\rangle)|0\rangle\nonumber\\
&+&\beta(\lambda_{}(|1010\rangle+|1111\rangle)|1\rangle\nonumber\\
\end{eqnarray}{}
\noindent The above state can be rearranged as:
\begin{eqnarray}    |\psi_{in,A,B}^{PFC,II}\rangle&=& \eta_1\lambda_{}(\alpha|0\rangle+\beta|1\rangle)+\eta_2\lambda_{}(\alpha|0\rangle-\beta|1\rangle)\nonumber\\    &+&\eta_1(\alpha|0\rangle+\beta|1\rangle)+\eta_2(\alpha|1\rangle-\beta|0\rangle)\nonumber\\
    &+&\eta_3\lambda_{}(\alpha|1\rangle+\beta|0\rangle)+\eta_4\lambda_{}(\alpha|1\rangle-\beta|0\rangle)\nonumber\\     &+&\eta_3(\alpha|1\rangle+\beta|0\rangle)+\eta_4(\alpha|1\rangle-\beta|0\rangle)
\end{eqnarray}
\noindent where,
\begin{eqnarray}
    \eta_{1,2}&=&=\frac{1}{2}\left[\left(|0000\rangle+|0101\rangle\right)\pm\left(|1010\rangle+|1111\rangle \right)\right],~\text{and,}\nonumber\\
    \eta_{3,4}&=&=\frac{1}{2}\left[\left(|0010\rangle+|0111\rangle\right)\pm\left(|1000\rangle+|1101\rangle \right)\right]
\end{eqnarray}
Bringing in the input qubit, Alice again uses measurements from Table~(\ref{ADC_SS_table}), and accordingly, Bob applies the unitary operations. The fidelity for this case is shown in Fig.~(\ref{PFC_newWM}). {Compared with the results for WM Protocol-I shown in Fig.~(\ref{PFC_sHWM}), the protected teleportation protocol employing WM Protocol-II also exhibits enhanced teleportation fidelity over a broad range of parameter values. This demonstrates that the improvement achieved by WM is robust against PFC noise and is not limited to the specific measurement protocol-I.}

\section{Success Probability of the Protocol}

{

The teleportation protocol described here is probabilistic in nature, since both the WM post-selection and the quantum
measurement reversal post-selection succeeds only with a finite
probability. This joint probability, which we refer to as the WMFR success probability $\mathcal{P}$, is defined as the trace of the unnormalized post-protocol Bell state $\rho_{ab}^{final}$ before normalization,

\begin{eqnarray}
  \mathcal{P}(q,\,\omega,\,r)
    &=& \mathrm{Tr}(\rho_{ab}^{final}) \nonumber\\
    &=& \mathrm{Tr}\!\left(
        \sum_{i,\,j\,\in\,\{0,1\}}
        N_i F_i E_j F_i M_i\,
        \rho_{ab}\,
        M_i^{\dagger} F_i^{\dagger} E_j^{\dagger} F_i^{\dagger} N_i^{\dagger}
      \right),\nonumber\\
    \label{eq:P_I}   
\end{eqnarray}

\noindent where $M_i$,
$F_i$,
$E_j$, and
$N_i$ are defined in Sec.~\ref{secii}, and $\rho_{ab} = |\psi_0\rangle\langle\psi_0|$
is the initial four-qubit Bell state of Eq.~(\ref{initial_state}). Evaluating
Eq.~(\ref{eq:P_I}) explicitly for each noise model yields the following results.




\begin{figure*}[!tbp]
  \centering
  \begin{subfigure}[b]{0.4\textwidth}
    \includegraphics[width=1\linewidth]{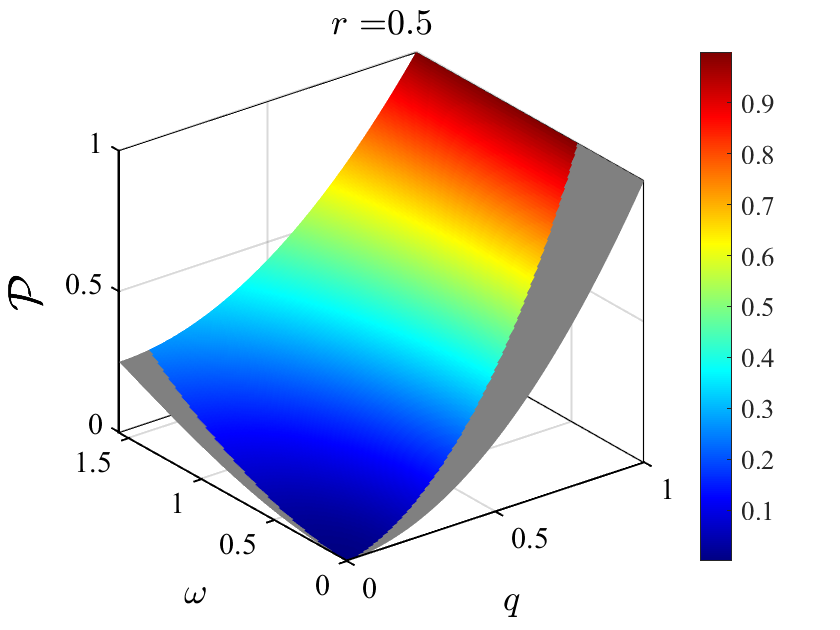}
    \caption{}
    \label{ADC_succProb_PI0p5}
  \end{subfigure}
  \hfill
  \begin{subfigure}[b]{0.4\textwidth}
    \includegraphics[width=1\linewidth]{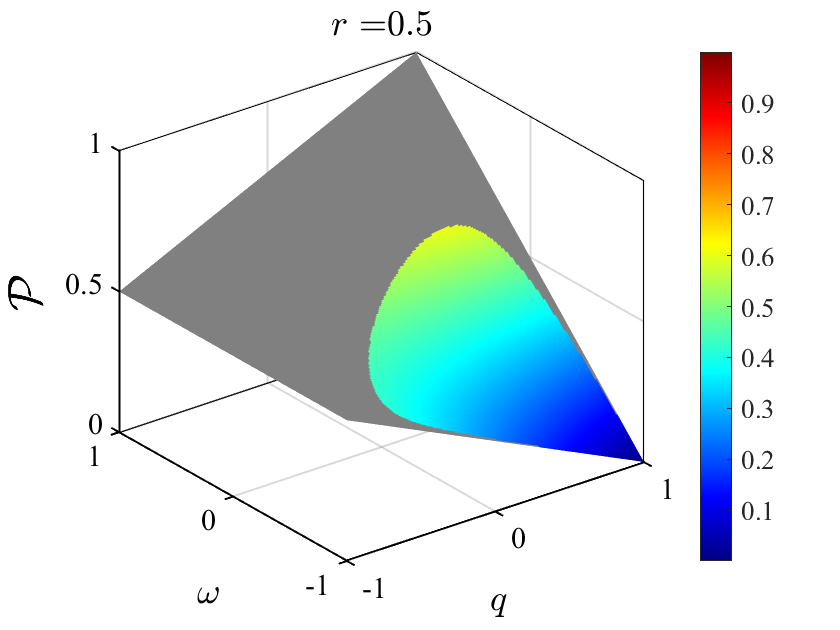}
    \caption{}
    \label{ADC_succProb_PII0p5}
  \end{subfigure}

  \vspace{0.5cm}

  \begin{subfigure}[b]{0.4\textwidth}
    \includegraphics[width=1\linewidth]{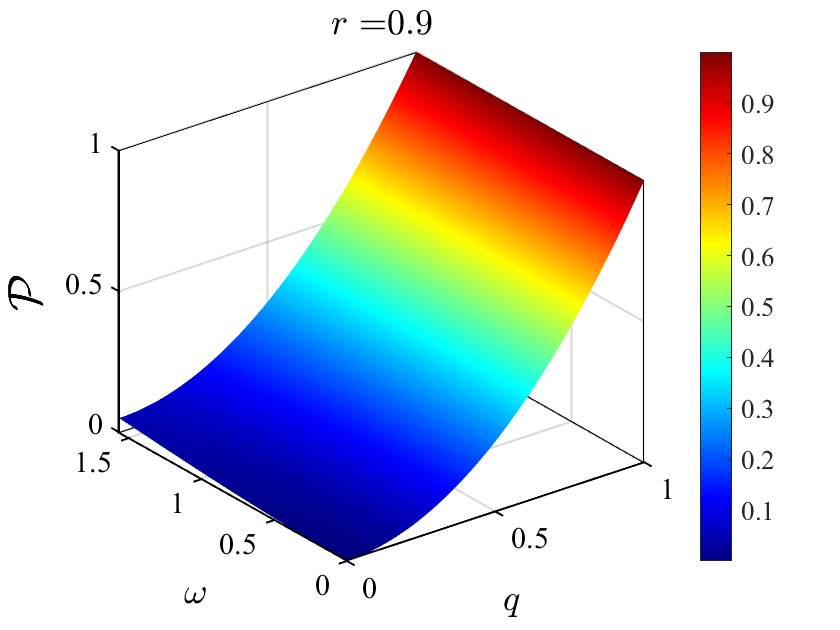}
    \caption{}
    \label{ADC_succProb_PI0p9}
  \end{subfigure}
  \hfill
  \begin{subfigure}[b]{0.4\textwidth}
    \includegraphics[width=1\linewidth]{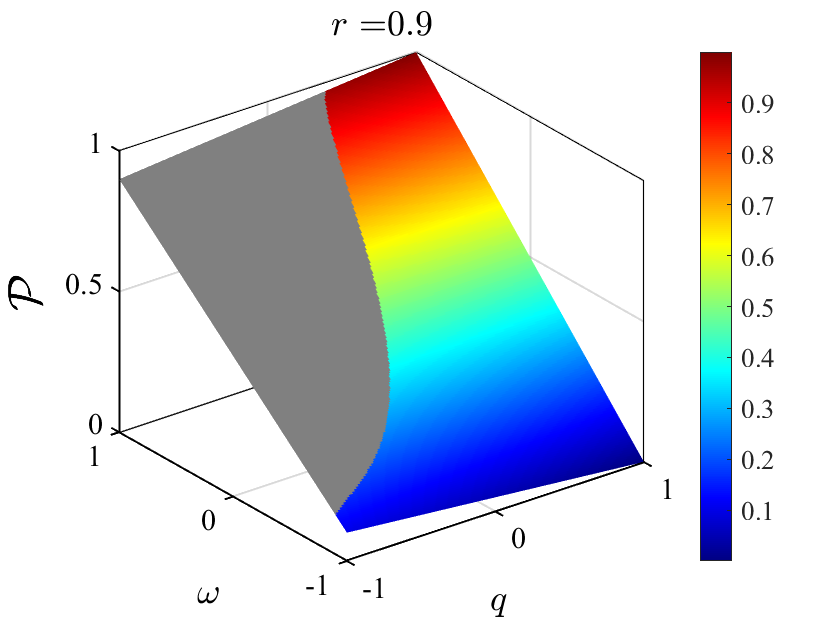}
    \caption{}
    \label{ADC_succProb_PII0p9}
  \end{subfigure}

  \caption{{Success probability $\mathcal{P}$ for the ADC under Protocol~I (Sec.~(\ref{sec_wmi})) is presented in sub-figures (a) for $r=0.5$, (c) for the case of $r=0.9$. For comparison, the success probability for WM Protocol-II (Sec.~\ref{sec_wmii}) is shown in sub-figures (b) and (d) for $r=0.5$ and $r=0.9$, respectively. The colored surface represents parameter combinations where $\mathcal{P}$
  corresponds to a fidelity improvement over the unprotected channel, while
  the gray region identifies combinations where no fidelity advantage is
  achieved. Protocol~I provides a fidelity benefit across a significantly
  broader region of the $(\omega, q)$ parameter space compared to
  Protocol~II in the case of ADC noise for both values of $r$, with the gray region disappearing
  entirely at $r=0.9$ for Protocol~I but persisting for Protocol~II.}}
  \label{ADC_succProb_comparison}
\end{figure*}
\subsection{ADC}
The success probability surfaces for the ADC under Protocol-I is shown in
Fig.~(\ref{ADC_succProb_PI0p5}) and Fig.~(\ref{ADC_succProb_PI0p5}), revealing a clear and physically meaningful
correspondence with the teleportation fidelity results of
Fig.~\ref{ADC_sHWM}. At $r = 0.5$, the coloured region of the
success probability surface representing parameter combinations
$(\omega, q)$ where $\mathcal{P}$
corresponds to a fidelity improvement and covers the majority of the
parameter space, with only a narrow grey strip appearing at large $q$
and small $\omega$. This strip coincides precisely with the region in
Fig.~(\ref{ADC_sHWM}) where the fidelity surface touches the grey
reference plane at small $\omega$ the weak measurement barely
pre-selects the damped component, so no coherence is recovered
regardless of how weak the quantum measurement reversal is set. It is to be noted that at $r = 0.9$ the grey area
disappears entirely from the success probability surface, confirming
that for strongly damped channels, Protocol-I provides a fidelity
advantage across the whole $(\omega, q)$ parameter space. This
is consistent with Fig.~(\ref{ADC_sHWM}) at $r = 0.9$, where the
fidelity surface rises well above the grey plane over a broad region,
with the peak fidelity approaching unity at intermediate $\omega$ and
small $q$. The key physical tradeoff is visible by reading both plots
together. The highest fidelity values in Fig.~\ref{ADC_sHWM}
(deep-red region, $\mathcal{F} \to 1$) correspond to the lowest
success probability values in Fig.~(\ref{ADC_succProb_PI0p5}) and Fig.~(\ref{ADC_succProb_PI0p5}), (deep-blue
region, $\mathcal{P} \to 0$), since aggressive WM pre-selection and strong quantum measurement reversal filtering retain only the most coherent fraction of events.

Comparing the ADC success probability results of Protocol-II
(Fig.~(\ref{ADC_succProb_PII0p5}), and Fig.~(\ref{ADC_succProb_PII0p9})) against Protocol-I (Fig.~(\ref{ADC_succProb_PI0p5}) and Fig.~(\ref{ADC_succProb_PI0p5}))
and their corresponding fidelity plots (Figs.~(\ref{ADC_newWM})
and ~(\ref{ADC_sHWM})) reveals a fundamental performance difference
between the two protocols for amplitude damping noise. At $r = 0.5$, the coloured (fidelity-improving) region of Protocol-II is confined
to a compact elliptical island centred near $(K_1, K_2) \approx (0, 0)$,
surrounded by a dominant grey surface. In contrast, Protocol-I
displays a coloured region covering nearly the entire $(\omega, q)$
plane at the same $r$. This contrast directly explains the fidelity
comparison in Figs.~(\ref{ADC_sHWM}) and Fig.~(\ref{ADC_newWM}), while
Protocol-I achieves $\mathcal{F}_{\max} \approx 0.999$ at $r = 0.5$,
Protocol-II still reaches a lesser value for a large parameter space, because the
narrower beneficial region in the $(K_1, K_2)$ space limits the
accessible fidelity improvement. At $r = 0.9$, the coloured region
of Protocol-II expands into a broader stripe but still occupies a
much smaller fraction of the total parameter space compared to
Protocol-I, for which the entire surface is beneficial. Whereas Protocol-I sustains higher fidelity compared to unprotected fidelity across the full
range of $r$, as confirmed by the maximum fidelity profile of
Fig.~(\ref{ADC_succProb_comparison}). 
The success probability surfaces thus not only quantify the filtering
cost of each protocol, but also provide a diagnostic of why
Protocol-I dominates Protocol-II specifically for amplitude damping
noise, while Protocol-II's symmetric structure becomes advantageous
for BFC as discussed below.

\subsection{BFC}

The success probability surfaces for the BFC under
Protocol-I are shown in Fig.~\ref{BFC_succProb_comparison}(a)
and Fig.~\ref{BFC_succProb_comparison}(c). These results
are consistent with the fidelity behaviour reported in
Fig.~\ref{BFC_sHWM}. For $r=0.5$ and $r=0.9$, the $\mathcal{P}$ remains in grey colour, representing no advantage of Protocol-I in the case of BFC, consistent with the fidelity results of Fig.~(\ref{BFC_sHWM}). 
Physically, the bit-flip channel
randomly exchanges the computational basis states
$|0\rangle$ and $|1\rangle$, thereby reducing the effectiveness
of the asymmetric WM-WMR structure employed in
Protocol-I. Consequently, although successful events can
still occur, the probability of obtaining a simultaneous fidelity enhancement becomes increasingly limited.
\begin{figure*}[!tbp]
  \centering
  \begin{subfigure}[b]{0.4\textwidth}
    \includegraphics[width=1\linewidth]{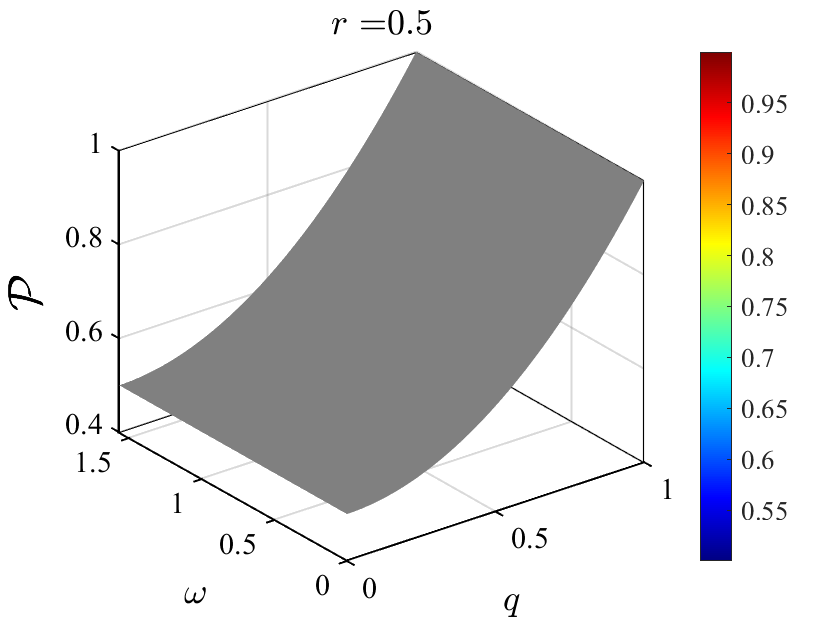}
    \caption{}
    \label{BFC_succProb_PI0p5}
  \end{subfigure}
  \hfill
  \begin{subfigure}[b]{0.4\textwidth}
    \includegraphics[width=1\linewidth]{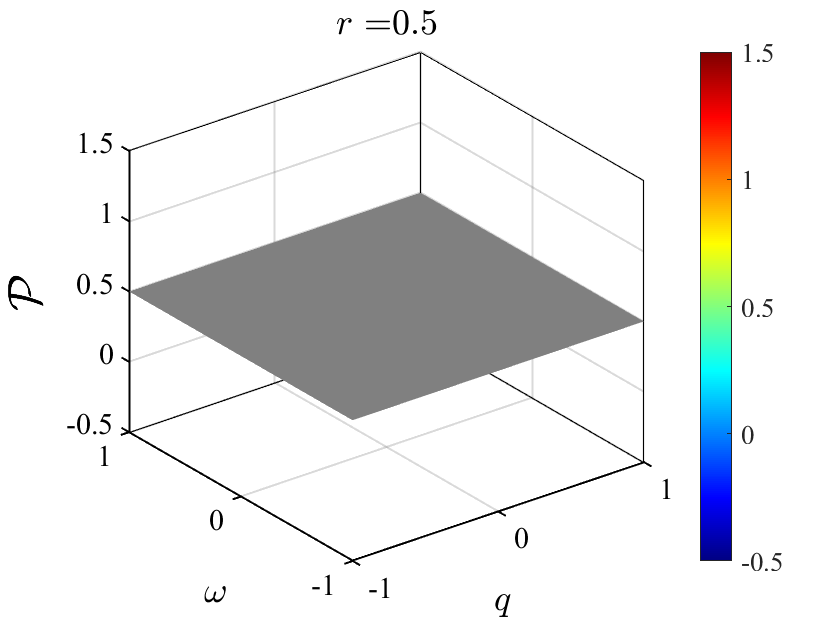}
    \caption{}
    \label{BFC_succProb_PII0p5}
  \end{subfigure}

  \vspace{0.5cm}

  \begin{subfigure}[b]{0.4\textwidth}
    \includegraphics[width=1\linewidth]{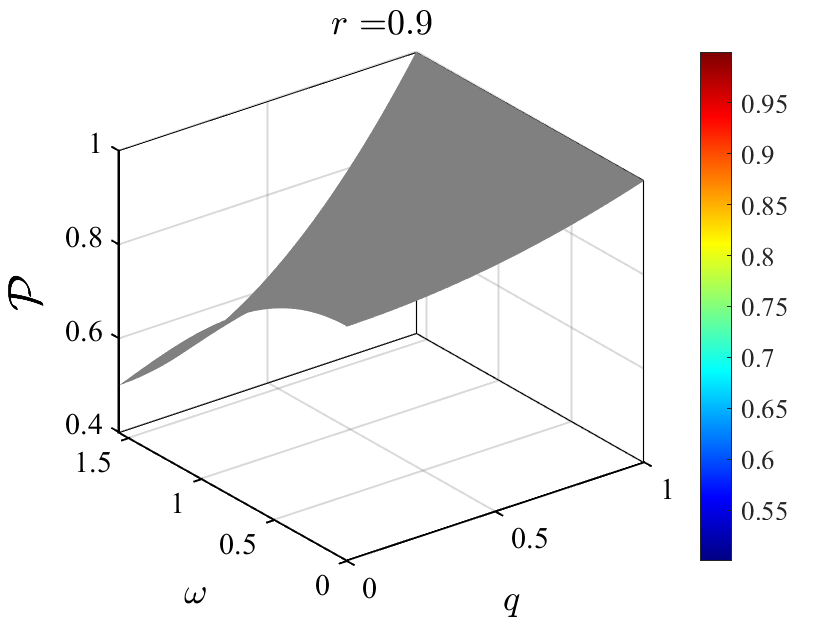}
    \caption{}
    \label{BFC_succProb_PI0p9}
  \end{subfigure}
  \hfill
  \begin{subfigure}[b]{0.4\textwidth}
    \includegraphics[width=1\linewidth]{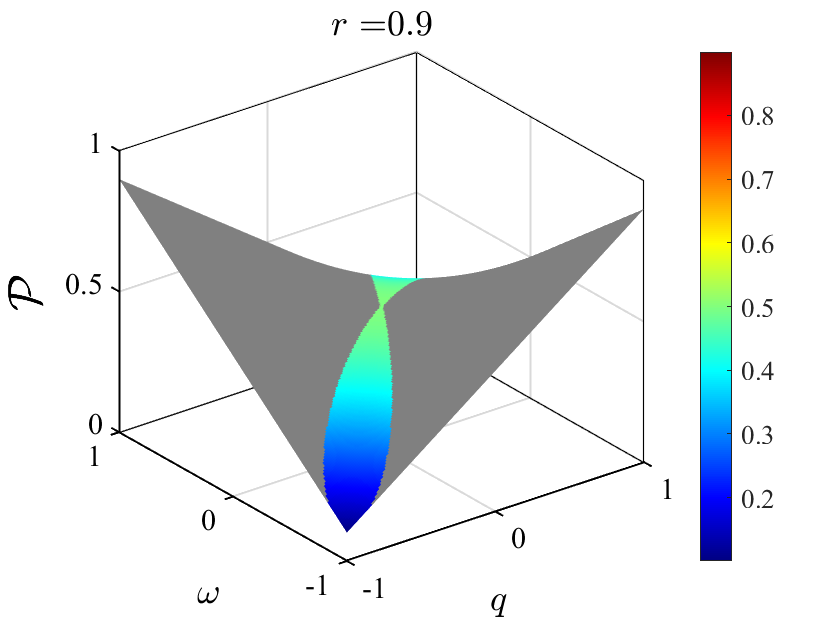}
    \caption{}
    \label{BFC_succProb_PII0p9}
  \end{subfigure}

  \caption{{Success probability P for the BFC under WM Protocol-I [Sec.~\ref{sec_wmi}] and WM Protocol-II [Sec.~\ref{sec_wmii}]. Sub-figures (a) and (b) show the results for $r=0.5$, while sub-figures (c) and (d) correspond to $r=0.9$. Within each row, the results for WM Protocol-I (left) are compared with those for WM Protocol-II (right). The colored surfaces denote the parameter regions where the corresponding protocol yields an improvement in teleportation fidelity over the unprotected channel, whereas the gray regions indicate parameter combinations for which no fidelity enhancement is obtained.
 }}
  \label{BFC_succProb_comparison}
\end{figure*}
In contrast, the BFC success probability surfaces for
Protocol-II shown in Fig.~(\ref{BFC_succProb_comparison}(b))
and Fig.~(\ref{BFC_succProb_comparison}(d)) reveal a markedly
different behaviour. At $r=0.5$, the coloured region is
essentially absent, and the success probability surface is
nearly flat, reflecting the observation in Fig.~(\ref{BFC_newWM}) that
Protocol-II provides little advantage over the
unprotected protocol at moderate bit-flip rates. However,
for $r=0.9$ a distinct coloured region emerges around a
narrow range of $(K_1,K_2)$ values, indicating that the appropriate tuning of the WM and WMR strengths can
simultaneously yield both a nonzero success probability and
improved teleportation fidelity. This behaviour directly
correlates with the fidelity enhancement observed in
Fig.~(\ref{BFC_newWM}), where Protocol-II outperforms Protocol-I at high
bit-flip decoherence. The symmetric structure of the
Protocol-II measurement and reversal operators appear
to be better suited to compensating the basis-state
exchange induced by the BFC, allowing a subset of
post-selected events to retain stronger quantum
correlations. Therefore, while Protocol-I remains more
effective for ADC, the success probability analysis
confirms that Protocol-II becomes the preferred strategy
for protecting teleportation against strong bit-flip noise.

\subsection{PFC}

The success probability surfaces for the PFC under
Protocol-I are shown in Fig.~\ref{PFC_succProb_comparison}(a)
and Fig.~\ref{PFC_succProb_comparison}(c). These results
closely follow the fidelity behaviour reported in Fig.~(\ref{PFC_sHWM}).
For $r=0.5$, the entire success probability surface is
grey, implying none of the corresponding parameter combinations provide a teleportation fidelity advantage
over the unprotected channel. This observation is in
complete agreement with Fig.~(\ref{PFC_sHWM}), where the fidelity
surface coincides with the unprotected teleportation
fidelity across the whole $(\omega,q)$ parameter space.
Physically, for moderate phase-flip noise, the asymmetric
weak measurement and reversal operations of
Protocol-I do not effectively compensate the phase
errors introduced by the channel.
\begin{figure*}[!tbp]
  \centering
  \begin{subfigure}[b]{0.4\textwidth}
    \includegraphics[width=1.2\linewidth]{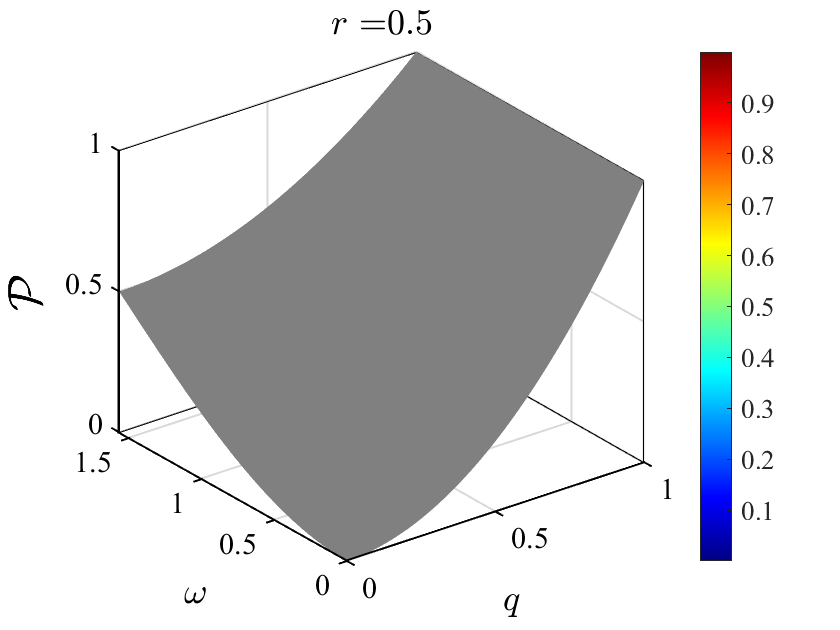}
    \caption{}
    \label{PFC_succProb_PI0p5}
  \end{subfigure}
  \hfill
  \begin{subfigure}[b]{0.4\textwidth}
    \includegraphics[width=1.2\linewidth]{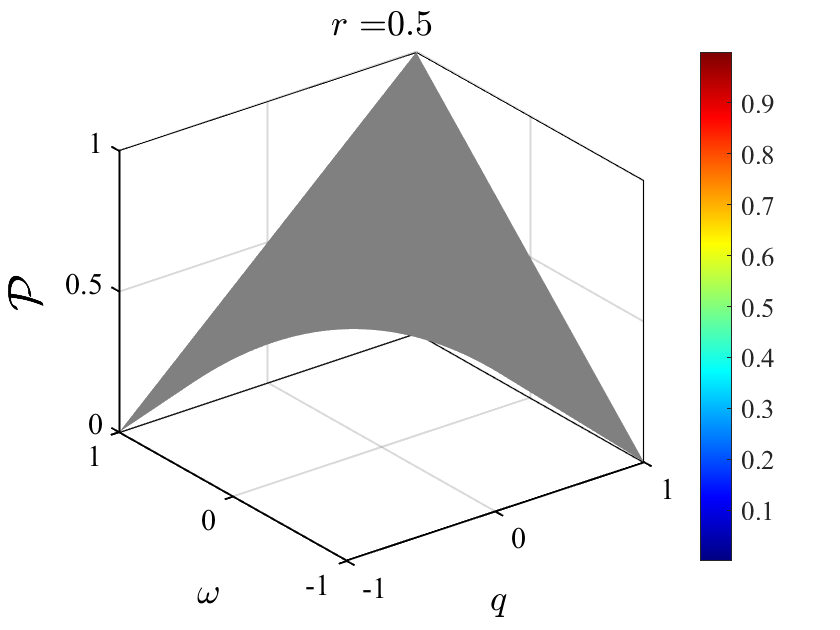}
    \caption{}
    \label{PFC_succProb_PII0p5}
  \end{subfigure}

  \vspace{0.5cm}

  \begin{subfigure}[b]{0.4\textwidth}
    \includegraphics[width=1.2\linewidth]{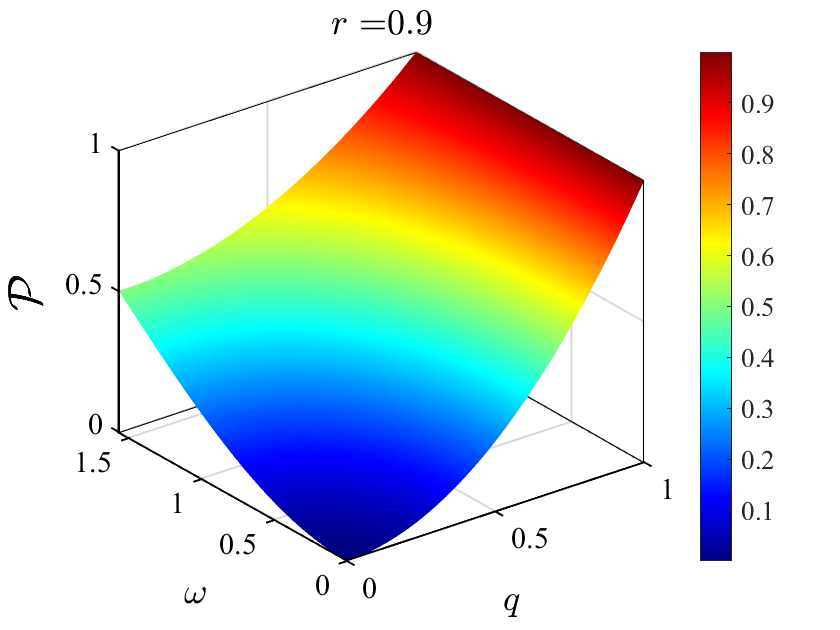}
    \caption{}
    \label{PFC_succProb_PI0p9}
  \end{subfigure}
  \hfill
  \begin{subfigure}[b]{0.4\textwidth}
    \includegraphics[width=1.2\linewidth]{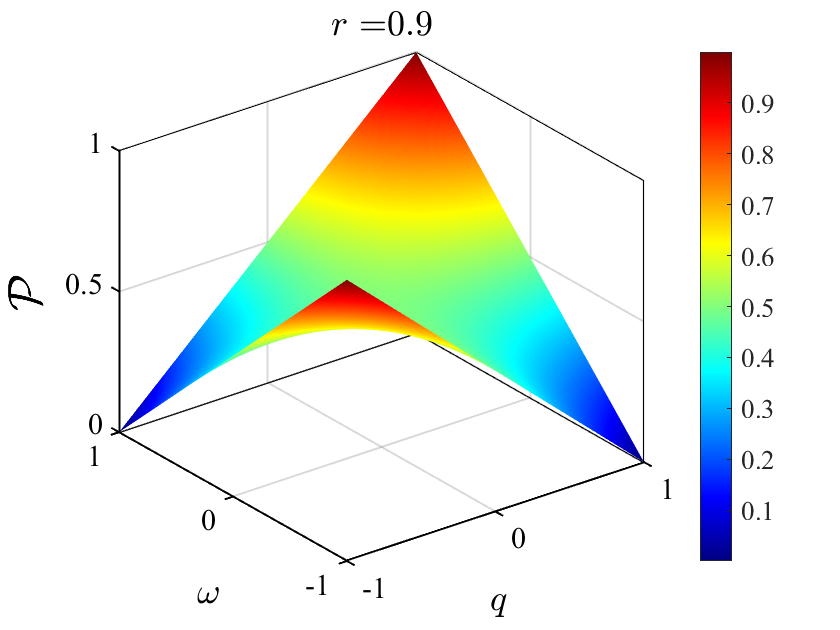}
    \caption{}
    \label{PFC_succProb_PII0p9}
  \end{subfigure}

  \caption{{Success probability $\mathcal{P}$ for the PFC under WM Protocol-I for decoherence rate (a) $r=0.5$, (c) $r=0.9$ and WM Protocol-II for (b) $r=0.5$ and (d) $r=0.9$. The colored surfaces represent the parameter regions where the corresponding protocol enhances the teleportation fidelity relative to the unprotected channel, while the gray regions denote parameter combinations for which no fidelity advantage is obtained. The absence of a colored region for $r=0.5$ in both protocols indicates that the WM-WMR scheme is ineffective in enhancing the teleportation fidelity under weak phase-flip decoherence.}}
  \label{PFC_succProb_comparison}
\end{figure*}
A different behavior emerges at $r=0.9$, where a
substantial coloured region appears in
Fig.~\ref{PFC_succProb_comparison}(c). The appearance
of this region indicates that suitable choices of the weak
measurement strength $\omega$ and reversal parameter
$q$ yield both nonzero success probability and a
teleportation fidelity exceeding that of the unprotected
channel. This result is consistent with Fig.~(\ref{PFC_sHWM}), where the
fidelity surface rises above the reference plane for a
significant portion of the parameter space. Similar to the
ADC case, the highest fidelities are obtained at the cost
of a reduced success probability, reflecting the usual
trade-off between state protection and post-selection
efficiency.

The corresponding success probability surfaces for
Protocol-II are displayed in
Fig.~\ref{PFC_succProb_comparison}(b) and
Fig.~\ref{PFC_succProb_comparison}(d). At $r=0.5$, the
entire surface remains grey, confirming that
Protocol-II, like Protocol-I, provides essentially no
fidelity improvement for moderate phase-flip noise.
This behaviour is fully consistent with the fidelity
results shown in Fig.~(\ref{PFC_newWM}), where the teleportation
fidelity remains equal to that of the unprotected
protocol throughout the $(K_1,K_2)$ parameter space.

However, at $r=0.9$, Protocol-II exhibits a significant
change. As seen in
Fig.~\ref{PFC_succProb_comparison}(d), the coloured
region expands over nearly the entire parameter space,
indicating that a broad range of weak measurement and
reversal strengths simultaneously provide nonzero
success probability and improved teleportation fidelity.
This observation directly correlates with the fidelity
enhancement reported in Fig.~(\ref{PFC_newWM}), where the fidelity rises
well above the unprotected value over an extensive
region of the $(K_1,K_2)$ plane. Compared with
Protocol-I, the beneficial region for Protocol-II is
considerably larger, demonstrating that the symmetric
structure of its WM-WMR operators is more effective in
counteracting strong phase-flip decoherence. Therefore,
the success probability analysis confirms that while both
protocols are ineffective against weak phase-flip noise,
Protocol-II becomes the more robust strategy in the
high-noise regime, offering fidelity enhancement across
a substantially broader parameter range.
}

\section{CONCLUSION}\label{conc}
\begin{figure*}[t!]
    \centering
    \begin{subfigure}[t]{0.32\textwidth}
        \centering
        \includegraphics[width=\textwidth]{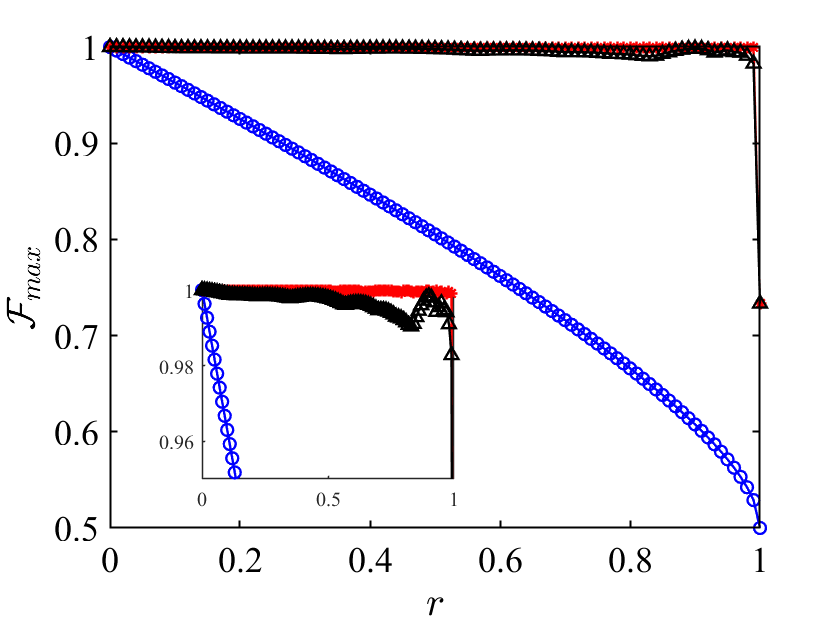}
        \caption{}\label{f10a}
    \end{subfigure}%
    \hfill
    \begin{subfigure}[t]{0.32\textwidth}
        \centering
        \includegraphics[width=\textwidth]{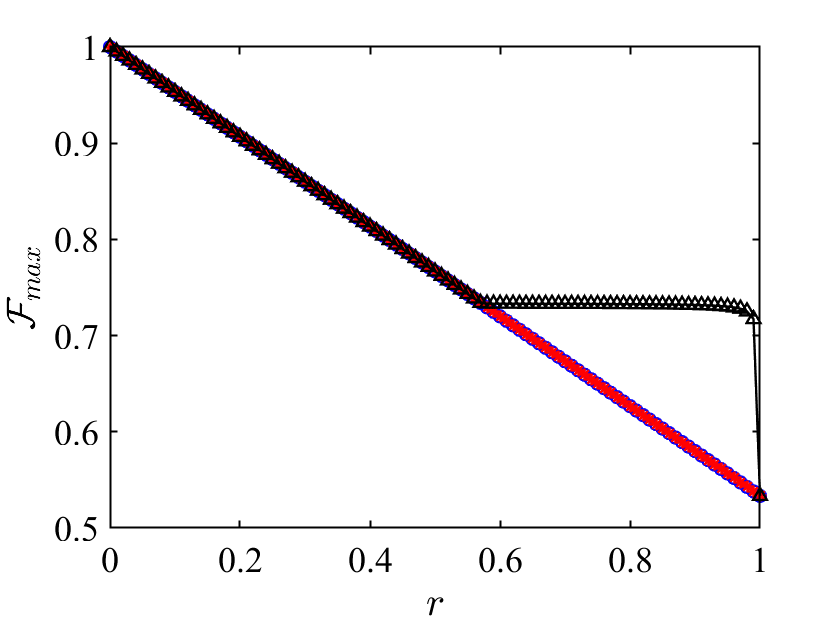}
        \caption{}\label{f10b}
    \end{subfigure}%
    \hfill
    \begin{subfigure}[t]{0.32\textwidth}
        \centering
        \includegraphics[width=\textwidth]{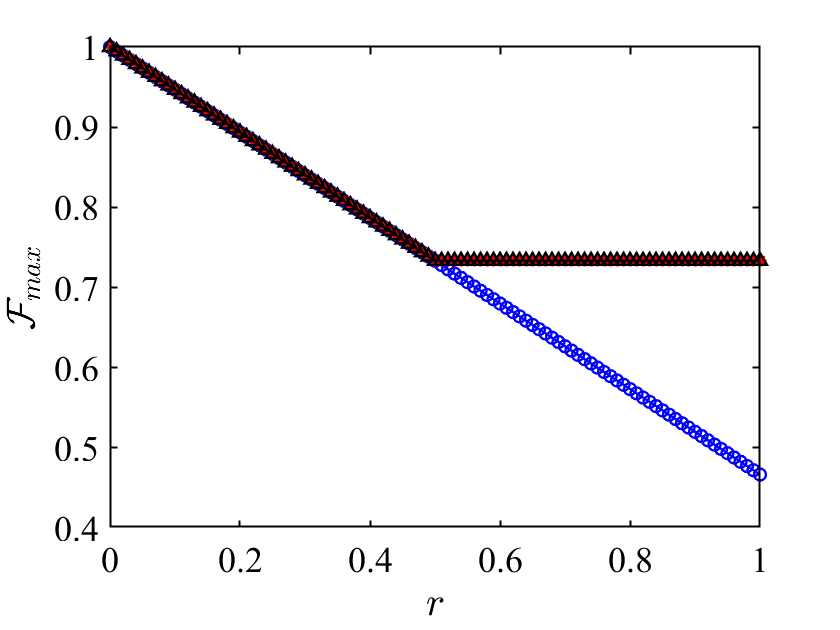}
        \caption{}\label{f10c}
    \end{subfigure}
    \caption{The plot for maximum fidelity versus decoherence for the case of (a) ADC, (b) BFC, and (c) PFC. The line with blue circles represents the maximum teleportation fidelity without any WM-WMR protocol, whereas the red star symbol's line represents WM-WMR protocol I, and the line with the black triangle symbol shows the behavior of maximum fidelity values using WM-WMR protocol II. {The inset in (a) shows a magnified view of the high-fidelity region $(0.95\leq\mathcal{F}\leq1)$ to clearly illustrate the variation in teleportation fidelity obtained using the two different WM-WMR protocols.}}\label{fmax}
\end{figure*}


In this work, starting from a 4-qubit entangled state, we presented a quantum teleportation scheme that enhances fidelity using weak measurement and reversal operations under three types of specific decoherence, namely, ADC, BFC, and PFC. We used the previously verified results of a 2-qubit entangled state teleportation under ADC decoherence using a specific weak-measurement protocol presented in \cite{harraz2022enhancing}. Building on their technique, here, we extend this idea and consider a teleportation protocol where the initial entangled state is considered to be a 4-qubit entangled state.  Furthermore, we analyzed two WM-WMR protocols, and their fidelities were calculated for not only ADC but also BFC and PFC decoherence channels as well. We started with Protocol I applied to all three noise channels: ADC, BFC and PFC. We observed that for ADC the maximum fidelity considering $r=0.5$ was $0.999$, and as $r$ is increased, the maximum value of the fidelity stayed close to $1$, as shown in Fig.~(\ref{ADC_sHWM}). Similarly, for BFC, from Fig.~(\ref{BFC_sHWM}) it is clear that the $\mathcal{F}$ value for $r=0.5$ is $0.7667$ and for $r=0.9$ the value of fidelity drops and the maximum fidelity is  $0.58$. Also, this maximum fidelity value is the same as that obtained for no WM protection at all. Therefore, it indicates that the protocol is not suitable for higher values of $r$ in the BFC case. Finally, for the PFC as shown in Fig.~(\ref{PFC_sHWM}) we got the maximum $\mathcal{F}$ values for $r=0.5$ as $0.733$ and for $r=0.9$ as  $0.734$ which are almost the same, reflecting a saturation of Fidelity after a certain value of $r$. However, it is to be noted that for the higher value of $r$, the fidelity without WM protection is significantly less than that compared to that with WM-WMR protection. Afterward, we introduced another protocol, the WM Protocol II, testing in the same manner for the three noises, ADC, BFC, and PFC. Starting from ADC we found for $r=0.5$ the maximum $\mathcal{F}$ value is $\mathcal{F}_{max}=0.81$ and for $r=0.9$ this value goes down to $0.754$ (see Fig.~(\ref{ADC_newWM})). This reflects the advantage of WM protocol-I over protocol II for this noise. For BFC, we found that for $r=0.5$ the $\mathcal{F}_{max}$ value is $~0.767$, whereas, as $r$ is increased to $0.9$ $\mathcal{F}_{max}=0.733$ (see Fig.~(\ref{BFC_newWM})). On comparison with protocol-I, this highlights that at higher values of $r$ protocol-II gives better performance than protocol-I when the noise is BFC. Concluding our results with PFC we found that for $r=0.5$ the $\mathcal{F}_{max}$ value is $0.733$ and for $r=0.9$ the maximum fidelity value still remains close to $0.733$ (see Fig.~(\ref{PFC_newWM})). Nevertheless, this $\mathcal{F}_{max}$ is still higher than that obtained without any protection. A complete profile of maximum fidelity with respect to $r$ is presented in Fig.~(\ref{fmax}). In all three cases, the $\mathcal{F}_{max}$ with protection can minimally be equal to that obtained without any protection. {In the case of ADC, Fig.~(\ref{f10a}), both protocols give maximum fidelity higher than that without any WM protection; however WM protocol-I shows a significantly better result in protection against decoherence. In contrast, for BFC in Fig.~(\ref{f10b}), WM protocol II gives higher teleportation fidelity when compared to WM protocol I as $r$ is increased. Therefore, for teleporting quantum information via BFC, applying WM Protocol II better protects the entangled state and improves fidelity. Whereas both the protection protocols are comparable when it comes to PFC as shown in Fig.~(\ref{f10c}).}

{
It is to be noted that the varying performance under different noise channels, namely ADC, BFC, and PFC, can be understood from their action on quantum states. ADC induces energy loss by damping $|1\rangle \rightarrow |0\rangle$, thereby directly reducing excitation probability and degrading entanglement. Weak measurement suppresses this decay by reducing the population of the excited state prior to transmission, allowing improved recovery via reversal operations.

In contrast, BFC probabilistically flips $|0\rangle \leftrightarrow |1\rangle$, which can be partially corrected by the pre- and post-flip operations used in the protocol. This explains the improved performance of Protocol-II at higher noise strengths due to the symmetric nature of WMR operators in Protocol-II when compared to Protocol-I.
PFC, however, introduces only phase errors without altering population. Since we define weak measurements such that it primarily targets only the amplitudes rather than phases, its effectiveness is limited in the case of PFC.}
Future research can build upon this work by extending the proposed framework to higher-dimensional quantum systems, such as qutrits and qudits, to explore its scalability and versatility in more complex state spaces.

\section{ACKNOWLEDGMENTS}
The authors acknowledge Sajede Harraz for her help on this work. SS acknowledges the financial support from the Department of Science and Technology (DST), Government of India, through the DST-INSPIRE Faculty Fellowship (DST/INSPIRE/04/2023/00184; Faculty Registration No. IFA23-PH303)

\appendix

\section{Comparison of WM protocol-I and WM protocol-II}\label{app1}
In this appendix, we compare the two protocols and briefly discuss why protocol-II works for ADC and BFC both. 
In protocol-I:
\begin{equation}
m_0 = 
\begin{pmatrix}
\cos(\omega/2) & 0 \\
0 & \sin(\omega/2)
\end{pmatrix}, \quad
m_1 = 
\begin{pmatrix}
\sin(\omega/2) & 0 \\
0 & \cos(\omega/2)
\end{pmatrix},\label{p1_m}
\end{equation}
\noindent with $0\leq\omega\leq\pi/2$ with reversal POVM operators 

\begin{equation}
n_0 = 
\begin{pmatrix}
q & 0 \\
0 & 1
\end{pmatrix}, \quad 
n_1 = 
\begin{pmatrix}
1 & 0 \\
0 & q
\end{pmatrix},\label{p1_n}
\end{equation}
\noindent where $0\leq q\leq1$.

And, the  protocol-II is,

\begin{equation}
m_0 = 
\begin{pmatrix}
K_1^+ & 0 \\
0 & K_1^-
\end{pmatrix}, \quad
m_1 = 
\begin{pmatrix}
K_1^- & 0 \\
0 & K_1^+
\end{pmatrix},\label{p2_m}
\end{equation}
where, $K_1^+=\sqrt{\frac{1+K_1}{2}}$, $K_1^-=\sqrt{\frac{1-K_1}{2}}$, and $K_1\in[0,1]$. With the reversal POVM operators:

\begin{equation}
n_0 = 
\begin{pmatrix}
K_2^+ & 0 \\
0 & K_2^-
\end{pmatrix}, \quad 
n_1 = 
\begin{pmatrix}
K_2^- & 0 \\
0 & K_2^+
\end{pmatrix},\label{p2_n}
\end{equation}
here, $K_2^+=\sqrt{\frac{1+K_2}{2}}$ and, $K_2^-=\sqrt{\frac{1-K_2}{2}}$, with $K_2\in[0,1]$.\\

Comparing Eq.~(\ref{p1_m}) and Eq.~(\ref{p2_m}), $K_1^+=\cos{\omega/2}$ and $K_1^-=\sin{\omega/2}$ with $K_{1/2}\in[-1,1]$. The idea is that $m_0$, $m_1$ in Protocol I (in Eq.~(\ref{p1_m})) and II (in Eq.~(\ref{p2_m})) are almost similar, just that Protocol II replaces the trigonometric parameter $\omega$ of Protocol I by the linearized parameter $K\in[-1,1]$. However, the difference lies in $n_0$, $n_1$. Protocol II (in Eq.~(\ref{p2_n})) keeps both amplitudes normalized symmetrically, whereas WMR operators in Protocol I (Eq.~(\ref{p1_n})) fix one diagonal entry to unity. This helps in ADC; however, for BFC, one needs a more symmetric measurement operator, and therefore, Protocol II provides better results in this case, because the symmetric parametrization allows better balancing between the $|0\rangle$ and $|1\rangle$ sectors.

Therefore, both protocols have some connection, i.e., Protocol I uses the trigonometric functions depending on $\omega$, and an asymmetric reversal parameter $q$. Whereas, Protocol II rewrites the same idea in a symmetric POVM form using parameters $K\in[-1,1]$. 
It is important to note that, in the case of the PFC, an effective protection scheme should employ POVM operators that directly target phase errors. However, the WM and WMR operators used in both Protocol I and Protocol II mainly modify the population amplitudes rather than their relative phases. Consequently, both protocols result in the same marginal improvement in teleportation fidelity under PFC.}

\bibliography{main_bib}

\end{document}